\documentclass[sigconf]{acmart}
\AtBeginDocument{%
  }

\setcopyright{acmlicensed}
\copyrightyear{2026}
\acmYear{2026}
\acmDOI{XXXXXXX.XXXXXXX}
\acmConference[CHI '26]{The ACM CHI conference on Human Factors in Computing Systems}{April 13--17,
  2018}{Barcelona, Spain}
\acmISBN{978-1-4503-XXXX-X/2026/04}

\usepackage{pifont}
\usepackage{tabularx}
\usepackage{multirow}
\usepackage{makecell}
\usepackage{booktabs,tabularx,makecell}
\usepackage{makecell}
\usepackage{enumitem} 

\usepackage{placeins}
\usepackage{float}
\usepackage{caption}
\usepackage{longtable}
\usepackage{makecell}
\usepackage{pdfpages} 
\usepackage{array}
\usepackage{ragged2e} 

\usepackage{xcolor}
\definecolor{accessibleBlue}{RGB}{0, 114, 178} 

\copyrightyear{2026}
\acmYear{2026}
\setcopyright{cc}
\setcctype{by}
\acmConference[CHI '26]{Proceedings of the 2026 CHI Conference on Human Factors in Computing Systems}{April 13--17, 2026}{Barcelona, Spain}
\acmBooktitle{Proceedings of the 2026 CHI Conference on Human Factors in Computing Systems (CHI '26), April 13--17, 2026, Barcelona, Spain}
\acmPrice{}
\acmDOI{10.1145/3772318.3791823}
\acmISBN{979-8-4007-2278-3/2026/04}

\begin{document}

\title{Exploring Teacher-Chatbot Interaction and Affect in Block-Based Programming}

\author{Bahare Riahi}
\email{briahi@ncsu.edu}
\orcid{0009-0005-4560-4857}
\affiliation{%
\department{Computer Science}
  \institution{North Carolina State University}
  \city{Raleigh}
  \state{NC}
  \country{USA}
}

\author{Ally Limke}
\orcid{0000-0002-4801-8723}
\author{Xiaoyi Tian}
\orcid{0000-0002-5045-0136}
\affiliation{%
\department{Computer Science}
  \institution{North Carolina State University}
  \city{Raleigh}
  \state{NC}
  \country{USA}
}

\author{Viktoriia Storozhevykh}
\orcid{0009-0002-3453-7781}
\author{Sayali Patukale}
\orcid{0000-0002-2960-9915}
\author{Tahreem Yasir}
\orcid{0000-0001-6505-1915}
\affiliation{%
\department{Computer Science}
  \institution{North Carolina State University}
  \city{Raleigh}
  \state{NC}
  \country{USA}
}

\author{Khushbu Singh}
\orcid{0009-0003-2803-0228}
\author{Jennifer Chiu}
\orcid{0000-0001-7663-5748}
\affiliation{
\department {School of Education and Human Development}
  \institution{University of Virginia}
  \city{Charlottesville}
  \state{VA}
  \country{USA}
}

\author{Nicholas lytle}
\email{nlytle3@gatech.edu}	
\orcid{0000-0001-7009-9905}
\affiliation{
\department {Computer Science}
  \institution{Georgia Institute of Technology}
  \city{Atlanta}
  \state{GA}
  \country{USA}
}

\author{Tiffany Barnes}
\email{tmbarnes@ncsu.edu}	
\orcid{0000-0002-6500-9976}
\author{Veronica Catet\'e}
\email{vmcatete@ncsu.edu}	
\orcid{0000-0002-7620-7708}
\affiliation{
\department {Computer Science}
  \institution{North Carolina State University}
  \city{Raleigh}
  \state{NC}
  \country{USA}
}

\renewcommand{\shortauthors}{Riahi et al.}

\begin{abstract}
AI-based chatbots have the potential to accelerate learning and teaching, but may also have counterproductive consequences without thoughtful design and scaffolding. To better understand teachers’ perspectives on large language model (LLM) based chatbots, we conducted a study with 11 teams of middle-school teachers using chatbots for a science and computational thinking activity within a block-based programming environment. Based on a qualitative analysis of audio transcripts and chatbot interactions, we propose three profiles: explorer, frustrated, and mixed that reflect diverse scaffolding needs. In their discussions, we found that teachers perceived chatbot benefits such as building prompting skills and self confidence alongside risks including potential declines in learning and critical thinking. Key design recommendations include scaffolding the introduction to chatbots, facilitating teacher control of chatbot features, and suggesting when and how chatbots should be used. Our contribution informs the design of chatbots to support teachers and learners in middle school coding activities.
\end{abstract}



\maketitle

\section{Introduction}
While integrating Artificial Intelligence (AI) applications in K-12 education has had a long and evolving history, the introduction of ChatGPT and other Large Language Model (LLM) powered technologies in 2022 accelerated the adoption of these tools \cite{gao2025exploring, kasneci2023chatgpt}. This rapid proliferation has created a litany of pedagogical and ethical concerns about problematic AI-usage in the classroom, with many stemming from uncertainty about how much of graded material (e.g., homework, assignments, exam answers) was completed by these LLMs rather than by the students themselves \cite{harvey2025don}. Products designed to detect LLM-generated text are not without error \cite{wu2025survey}, and there are inconsistencies at the school, course, and teacher-level about what types of LLM interactions constitute plagiarism \cite{nguyen2023ai}. For instance, a student generating small sections of code or test cases for a larger coding project may or may not be seen as acceptable. Additionally, most of these AI-interactions are student-self directed and in environments outside of either the visibility or control of the teacher. 

This type of use and misuse of AI in classrooms can potentially have impacts on learning outcomes. While our understanding of these effects are still evolving, early evidence points to the idea that while unrestricted student use can lead to faster and more accurate completion of activities in the short term, there can be long-term learning consequences. Research points to the idea that offloading cognitive effort to an AI may reduce the mental struggle necessary for deep learning, particularly in domains like programming where "productive struggle" is a key component of skill acquisition \cite{chi2014icap}. When students rely on an LLM to generate code, they may bypass the critical thinking and problem-solving processes that lead to lasting understanding \cite{gerlich2025ai}.
Companies are responding to these challenges by creating LLM systems targeted towards more effective pedagogical practices (e.g.  ChatGPTs StudyMode \cite{openai2025studymode}). 
\textbf{Professional development workshops} will allow researchers and chatbot designers to capture the range of concerns teachers have about unwanted interactions with the chatbot as well as set standards for desired AI  behaviors for both student-AI and teacher-AI interactions. Additionally, the range of learning environments and activities where these chatbots can be employed and differences in teachers' pedagogical style suggest that a ‘one-size-fits-all’ system with no instructor control over behavior is unacceptable. These complexities highlight the need for psychological frameworks such as Technology Acceptance Models (TAM) \cite{davis1985technology} and  Theory of Planned Behavior (TPB) \cite{ajzen1985intentions} to explain teachers’ evaluations and adoption decisions, and we mapped our analysis to these frameworks to interpret how their emotions, perceptions, and sense of control shaped their intention to use the chatbot.

In the summer of 2025, we invited 25 middle school science teachers to participate in a professional development focused on integrating computational thinking (CT) and block-based programming activities into their science classrooms. As part of this training, we included a module in which teachers completed a lesson within a block-based environment that had an integrated ‘chatbot’ component. This chatbot had a range of features including the generation of code from written description as well as the ability to answer questions regarding programming concepts generally or questions related to how to use the block-based environment specifically. Our aim was to use data collected from both these learning modules and a set of follow up discussions with the teachers to guide our understanding of how teachers interact with these types of technologies in practice, and how teachers would want a more formally developed AI system to operate. This aim is characterized by the following research questions: 
 
Research Questions:
\begin{enumerate}
    \item RQ1. How does interacting with chatbots for learning programming impact middle school teachers’ affect and attitudes?
    \item RQ2. What are teachers’ perspectives about the benefits and risks of using an LLM for block-based programming in their classrooms for students and for teachers?
\end{enumerate}

\section{Related Work}
\subsection{How Teachers Interact with LLMs in Pedagogical Environments}
 Kim et al. (2024) conceptualized teacher AI interactions as cognitive, socio-emotional, and artifact-mediated \cite{kim2024leading}. Cognitive interactions involve teachers' reasoning about AI outputs, selecting or modifying tasks, and making instructional decisions. Socio-emotional interactions reflect the trust, confidence, and attitudes of teachers toward AI, influencing when and how they use AI suggestions. Artifact-mediated interactions occur when teachers engage iteratively with AI-generated outputs, treating them as artifacts to support lesson design, instructional decision-making, and student learning. This framework, applied to various AI tools and classroom dashboards, can also be observed in studies examining the use of LLMs, such as ChatGPT.

Holstein et al. (2018) evaluated Lumilo, a wearable mixed reality tool for K-12 teachers, providing real-time insights into student learning and behavior \cite{holstein2018classroom}. Teachers engaged cognitively by planning interventions, socio-emotionally by evaluating trust and usability, and artifact-mediated through learning analytics. Lumilo also highlighted challenges with LLM use, such as information overload, anxiety, and the need to balance interpretability with accuracy.

Across these studies, teachers’ interactions with LLMs and AI tools reflect cognitive, socio-emotional, and artifact-mediated dimensions. Teachers use AI outputs to reason, adapt, or reject; navigate trust in AI; and iteratively interact with AI-generated artifacts to shape instruction. They prioritize maintaining agency, seeking tools that reduce burdens while preserving pedagogical control and ethical responsibility \cite{yusuf2025towards, feldman2025impact}. Overall, effective teacher–LLM interactions are best seen as collaborative partnerships where AI augments, not replaces, teacher expertise.
LLMs and generative AI have quickly entered schools, generating excitement and concerns. Teachers view ChatGPT as a valuable tool for saving time and supporting computer science lesson planning \cite{powell2024}, but also recognize challenges like academic integrity, student privacy, AI reliability, school restrictions, and ChatGPT’s age requirement, which complicate classroom integration.

Teachers' confidence and competencies significantly impact their adoption of LLM tools \cite{zhou2020}. For instance, Reichert et al. described a workshop for secondary teachers on ChatGPT that improved understanding and increased positive attitudes from 45\% to 68\%  \cite{reichert2024}. A survey of 102 STEM teachers in Germany found that teacher competence was the strongest predictor of ChatGPT use, with future use driven by perceived teaching benefits. Concerns about privacy, copyright, and reliability had minimal impact, suggesting that professional development and competence-building shape teachers' expectations for LLM design and use \cite{beege2024}.

Key design considerations include supporting active learning, providing ethical guidance, facilitating prompt engineering, offering structured materials, overcoming school/age barriers, ensuring continuous support, and fostering critical thinking \cite{zawacki2019systematic}. By addressing misconceptions, offering structured guidance, and building teacher competencies, AI can become a transformative tool for educators.
Insights from a design-based research (DBR) study on human-centered AI in STEM classrooms highlight teachers' perspectives on AI's role in managing collaborative learning and scaffolding instruction \cite{roll2016evolution}. Teachers valued AI that could reduce routine workload, offer structured guidance, and enhance classroom management without compromising their professional judgment \cite{lin2021engaging}. Teachers expected AI to complement their expertise, provide timely support, and seamlessly integrate into workflows. Their expectations for autonomy, control, and practical support are crucial for designing AI tools that are both useful and adoptable \cite{andersen2022collaborative}.

Collectively, these findings suggest that teachers' perceptions, confidence, and expectations are critical for the successful adoption of AI \cite{celik2025teacher}. LLMs and other AI tools should be designed in accordance with human-centered AI principles, aligning with teachers’ pedagogical expertise, workflows, and classroom realities. Such tools should empower teachers by augmenting their professional judgment, providing structured guidance and timely support, and enabling creativity and active engagement, while maintaining human control and accountability in the learning process \cite{shneiderman2020,yang2021}.

\subsection{Design/Use of LLM in Secondary and Coding Classrooms }
In recent years, several professional integrated development environments (IDEs), such as GitHub Co-Pilot, Cursor, and Claude, have integrated LLM technology to assist developers. However, less attention has been given to IDEs for classrooms, which require additional design considerations. Secondary classrooms typically use block-based environments like Scratch \cite{resnick2009scratch,maloney2010scratch}, MIT app inventor \cite{patton2019app} and UC Berkeley's Snap! \cite{harvey2013snap} to introduce programming concepts. The type of LLM support needed for students learning programming differs from that for professional developers. As LLM-integrated IDEs are still new, it's unclear what support students need and when to provide it for optimal learning.

Recent studies on large language models (LLMs) in secondary coding classrooms have explored a range of support, from code generation to structured scaffolding.
Tools offering code adaptation and stepwise guidance, such as Parsons puzzles, minimal fix units, and next-step hints, yield stronger learning outcomes than tools that generate full solutions \cite{Hu2025From,Hou2024CodeTailor,Hou2025Personalized,Deng2025Evaluating}. These interactive approaches promote active problem solving, extending practice time, improving retention, and reducing over-reliance on automated answers \cite{riahi2025comparative}. 
Personalized scaffolding, tailored to address student errors or prior work, has shown to further enhance engagement and learning outcomes \cite{del2024Automating}. 
Feedback systems powered by LLMs can provide instant, precise, and individualized responses, sometimes outperforming human instructors in error detection \cite{Lee2025GPT,Scholz2025Partnering}. Yet, challenges persist, as LLM feedback can be misleading and lacks adaptability to classroom dynamics \cite{Roest2024Next,Hellas2023Exploring}.  

For teachers, the effectiveness of LLM-based tools depends on integration into pedagogy and professional development. Educators value systems that encourage reflection and deeper understanding rather than direct answer-giving, often supported by guardrails that prompt students to engage in self-correction \cite{Liffiton2023CodeHelp,dai2024effect}. At the same time, design tensions emerge between providing sufficient scaffolding and preserving learner agency. For example, in \textit{Cognimates Scratch Copilot} \cite{druga2025scratch}, students leveraged AI suggestions for ideation, debugging, and asset creation but actively adapted or rejected outputs to maintain creative control.
These findings highlight that the value of LLMs lies not in their raw generative capability but in their careful design as collaborative learning partners that balance efficiency, accuracy, and student autonomy. However, there is still limited understanding of teacher perspectives and perceptions regarding the use of LLMs in classrooms, with a noticeable gap in detailed evidence on how teachers experience these tools. Specifically, more research is needed to understand the challenges teachers encounter, their perceptions of the benefits and usefulness of AI tools, and the design features they prefer, as these factors strongly influence adoption decisions.

\section{Methods}
\subsection{Participants}
Our LLM chatbot study took place during a teacher professional development (PD) workshop on infusing computational thinking and block-based programming into middle grades science classrooms. The professional development had both experienced lead teachers (N=8) and novice participant teachers (N=17) (Table \ref{tab:commands}) . The lead teachers had previous experience with computing education and had taken part in professional development with our research group before. The participant teachers were newer to computing and hadn't been involved in our professional development programs. The lead teachers did the 1-hour study on day 0, during their pre-PD training, while the participant teachers did it on day 2, with the lead teachers facilitating the session.

\begin{table*}
  \caption{Demographics of Teachers in LLM study}
    \resizebox{0.7\textwidth}{!}{
  \begin{tabular}{cccc}
    \toprule
    Type & Gender & Subjects & Skills \\
    \midrule
    \texttt{Lead: n=8} & 8 female& 
    \makecell[c]{6 Sci, 1 Math\\ 1 CS, 1 ELA} & 
    \makecell[c] {8 Scratch+Snap!\\ 7 code.org }\\
    \\
    \texttt{ Participants:n=17}& 
    \makecell[c] {12 female \\ 5 male} 
    & \makecell[c] {10 Sci, 3 Math\\ 1 CS, 3 other }
    & \makecell[c] {11 Scratch\\ 3 code.org}\\
    \bottomrule
  \end{tabular}}
  \Description{Table 1: Participant Information
This table provides an overview of the participant demographics for the study. It is divided into two sections: Lead and Participants. For the Lead group, there are 8 female participants with a background in various subjects: 6 in science, 1 in math, 1 in computer science (CS), and 1 in English language arts (ELA). They have proficiency in programming tools like Scratch+Snap! (8 leads) and code.org (7 leads). In the Participants group, there are 17 individuals, including 12 females and 5 males. Their subject expertise includes 10 in science, 3 in math, 1 in CS, and 3 in other areas. Their programming skills include 11 participants with experience in Scratch, 3 with experience in code.org, and no additional tools or languages specified.}
  \label{tab:commands}
\end{table*}

\subsection{Chatbot-integrated programming environment: stax.fun)}

\begin{figure}[ht]
  \centering
  \includegraphics[width=0.8\columnwidth]{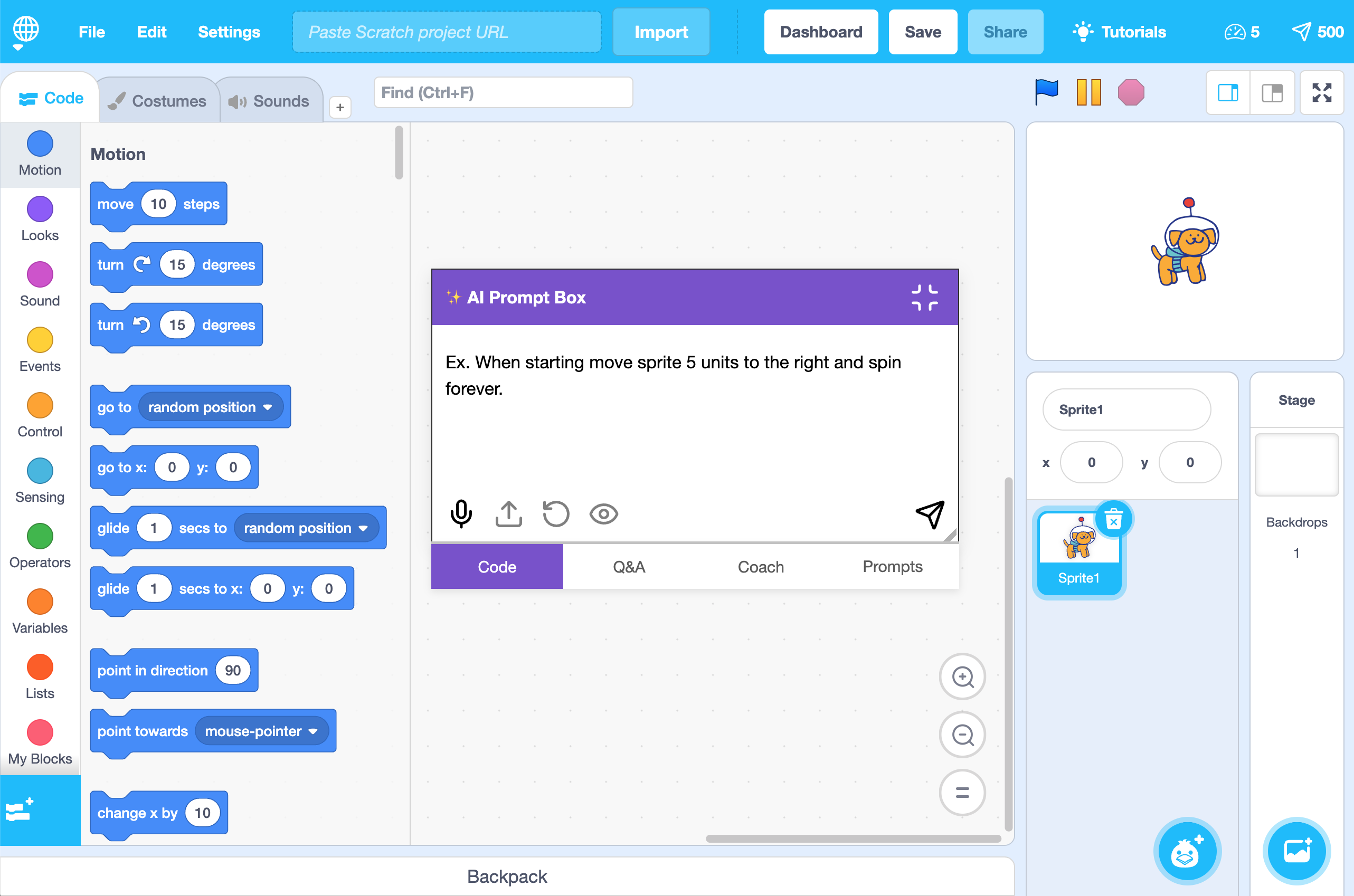}
  \caption{Stax.fun Platform overview and its chatbot tabs and features}
  \label{fig:stax}
  \Description{a four-tabbed modal window with different LLM options overlayed on the scratch programming environment}
\end{figure}

\begin{figure}[ht]
  \centering
  \includegraphics[width=0.8\columnwidth]{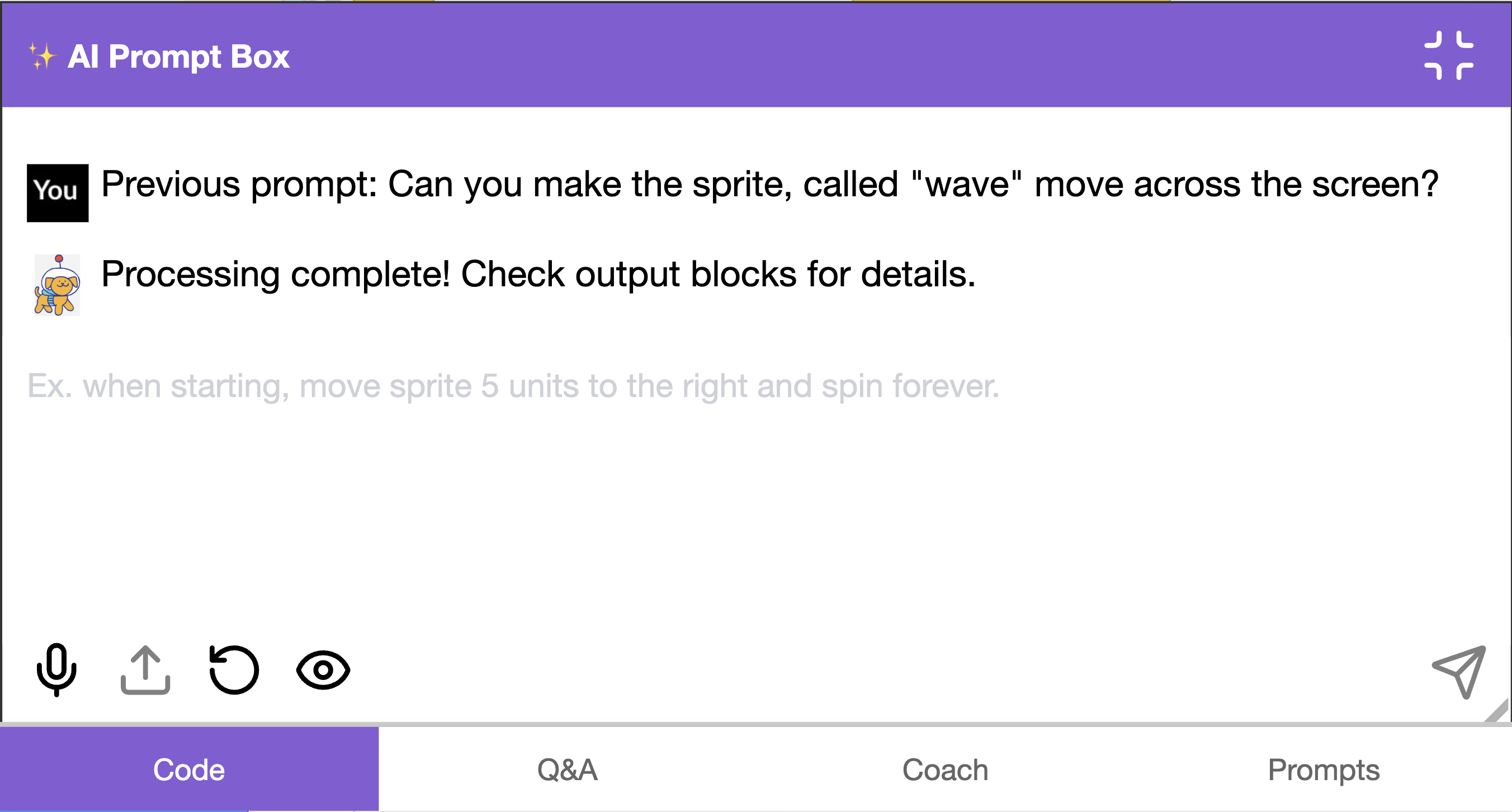}
  \caption{Code Tab generating code snippets, providing error corrections }
  \label{fig:code}
  \Description{generating code snippets, providing error corrections}
\end{figure}

\begin{figure}[ht]
  \centering
  \includegraphics[width=0.8\columnwidth]{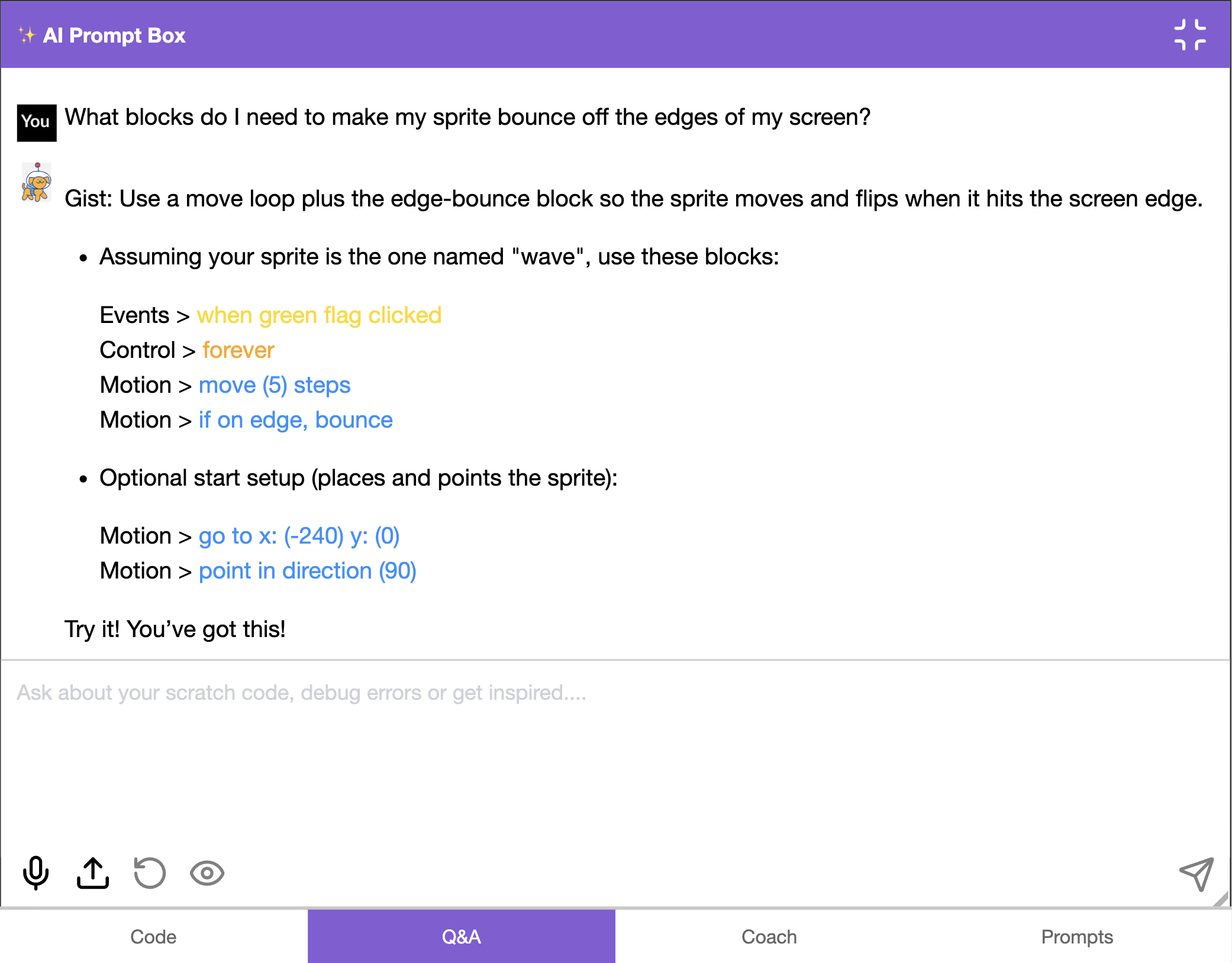}
  \caption{Q\&A Tab offering real-time answers, explanations, or clarifications }
  \label{fig:QandA}
  \Description{Q\&A Tab offering real-time answers, explanations, or clarifications }
\end{figure}

\begin{figure}[ht]
  \centering
  \includegraphics[width=0.8\columnwidth]{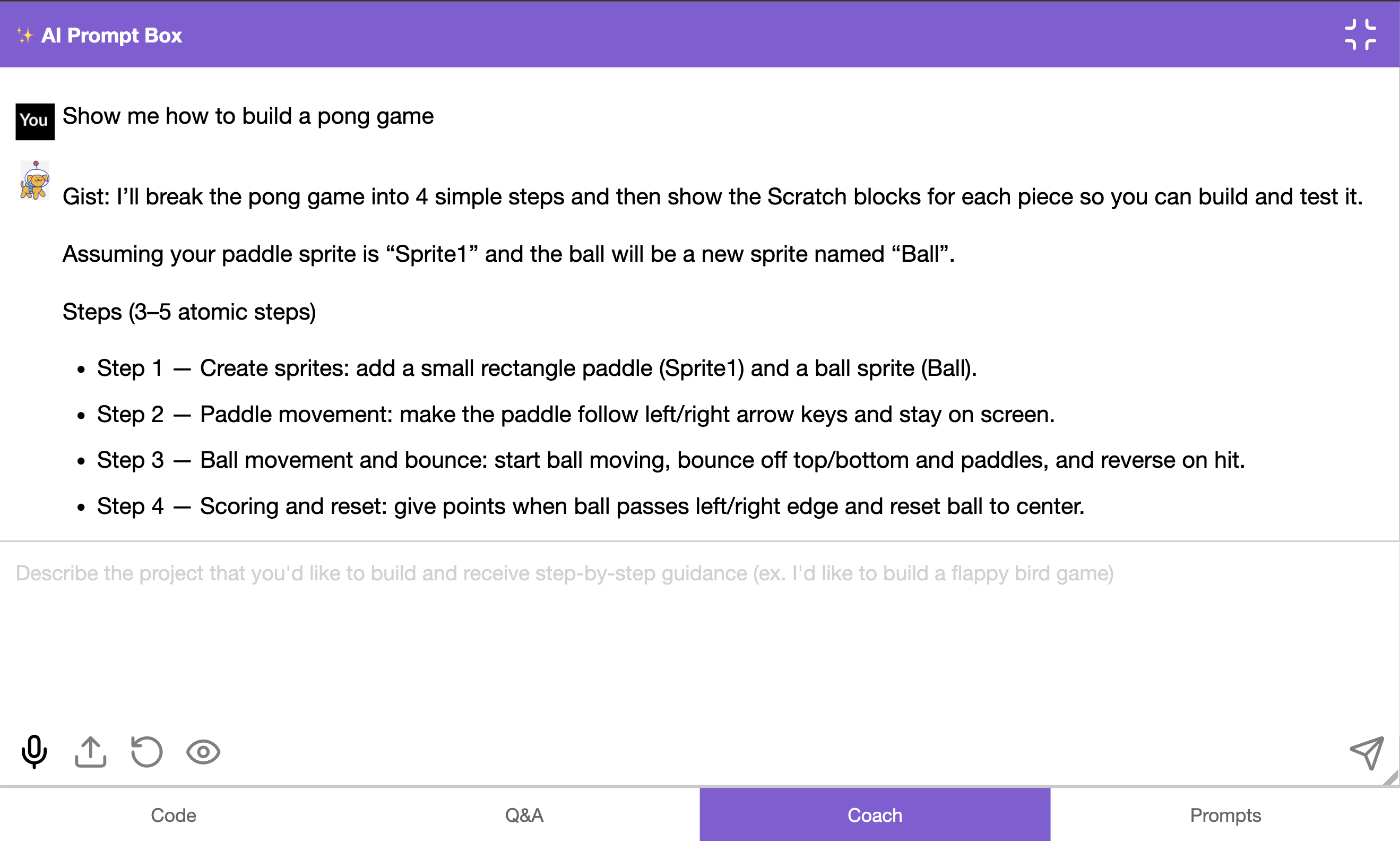}
  \caption{Coach Tab offering hints, tips, and suggestions }
  \label{fig:coach}
  \Description{Coach Tab provides step-by-step guidance, offering hints, tips, and suggestions}
\end{figure}

\begin{figure}[ht]
  \centering
  \includegraphics[width=0.8\columnwidth]{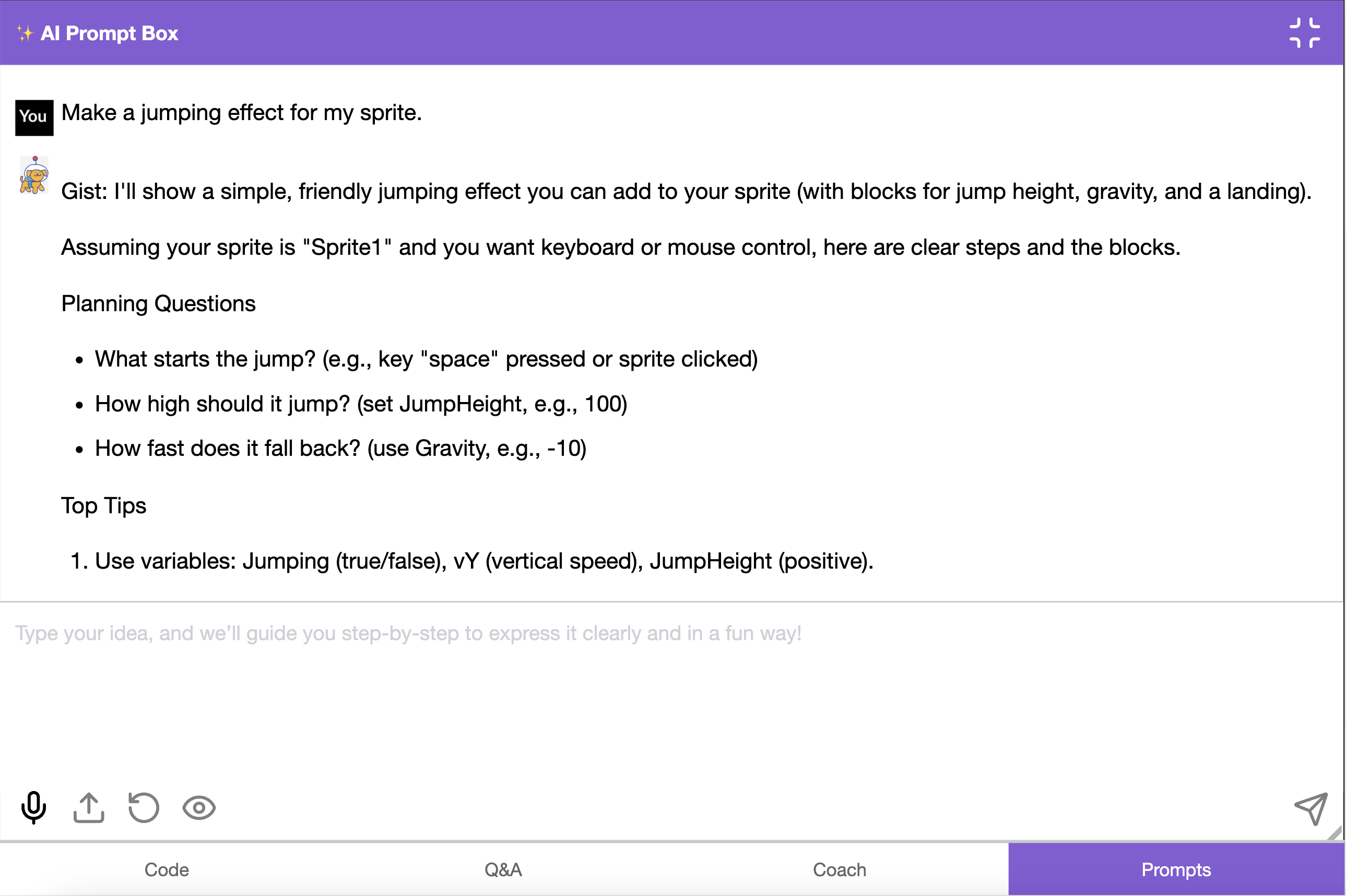}
  \caption{Prompt Tab refine and structure the prompt to align with the desired outcome}
  \label{fig:prompt}
  \Description{refine and structure the prompt to align with the desired outcome}
\end{figure}

The programming environment we chose to use for this study is \textit{Stax.fun (Figure \ref{fig:stax})} offers an AI copilot for block-based coders that can generate, debug, and troubleshoot Scratch programming code through four different prompt modes. The developers describe these modes as \textit{Code tab (Figure \ref{fig:code})}: generate visual blocks using natural language prompts;  \textit{Q\&A tab (Figure \ref{fig:QandA})} : get answers, insights, and creative ideas about visual block coding; \textit{Coach tab (Figure \ref{fig:coach})} : generate step-by-step guidance on building a project idea from scratch; \textit{Prompts tab (Figure \ref{fig:prompt})} : generate refined prompts based on the user's prompts for more accurate and creative results. Stax also supports importing and exporting native Scratch projects (.sb3 files). Each participant logged in using a unique student account linked to the researchers' teacher account with our starter code preloaded. 

\subsection{Study Activities and Tasks}
The study involved 11 key activities designed to engage participants with the chatbot, organized into two main phases: \textit{Getting Familiar with the Chatbot} (Steps 1-5) and \textit{Coding Wave Activities} (Steps 6-11). 
\begin{enumerate}
    \item Getting Familiar with the Chatbot’s interface and its various features (Steps 1-5):

    \begin{itemize}
        \item \textbf{Step 1-5:} Tasks such as understanding the functionality of the chatbot and familiarizing themselves with different tabs (Code, Coach, Q\&A, etc.).
    \end{itemize}
                                    
\item Coding Wave Activities (Wave interactions) (Steps 6-11):

    \begin{itemize}
        \item \textbf{Step 6 (Starter Code):} Initial task involving basic coding concepts.
        \item \textbf{Step 7 (Coding - Curtain):} Participants worked on coding tasks related to simulating the interaction of light waves with a curtain.
        \item \textbf{Step 8 (Coding - Glass):} Tasks focused on how light waves interact with glass.
        \item \textbf{Step 9 (Coding - Mirror):} Coding interaction between light waves and mirrors.
        \item \textbf{Step 10 (Sound Wave Activity):} Participants applied the same principles to sound waves, coding how sound interacts with various materials.
        \item \textbf{Step 11 (Plastic-bag Activity):} Focused on coding the interaction of light waves with a plastic bag, using a ghost effect to represent partial transmission.
    \end{itemize}
    \end{enumerate}
                                
\subsection{Procedures}
Our data collection was conducted on Day 2, while Day 1 was dedicated to introducing the participants to block-based programming.
\textit{Day 1 (Introductory Activity):} Participants were introduced to the block-based programming language Snap!, which would be used throughout the PD. They then completed an activity in Snap!, which was adapted from a middle school science standard on wave interactions, requiring participants to program the behavior of waves. Slides guided participants step-by-step through the activity and included answer checks. Participants had 1 hour to complete the activity and worked in pairs on their laptops.
This day was not part of the data collection procedure; it was only intended as an introduction for the participants.
\textit{Day 2 (Wave Activity):} We conducted a 30-minute think-aloud study with teachers in groups of 3-4. Lead teachers did not complete the introductory activity since they were already familiar with it and served as facilitators for the other teachers.
During the study, participants were introduced to a large language model (LLM)-powered chatbot for block-based programming in a Scratch IDE, stax.fun. Participants spent approximately 10 minutes learning about and experimenting with the chatbot. They then completed the same programming activity as on Day 1. To facilitate the use of the chatbot, the guided slides were modified to include prompts encouraging participants to engage with the chatbot. Participants were also encouraged to use the chatbot as needed to assist them with the programming tasks. Participants were asked to record their screens and audio using Zoom during this activity.

\begin{figure}[ht]
  \centering
  \includegraphics[width=\columnwidth]{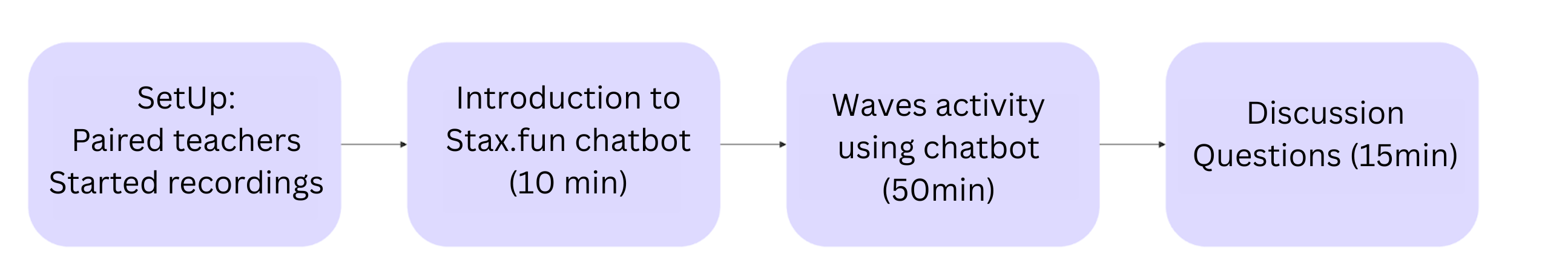}
  \caption{Process of the study with stax.fun chatbot}
  \label{fig:days}
  \Description{A four sequence flow chart: set-up, introduction (10 mins), chatbot activity(50 mins), discucssion (15mins)}
\end{figure}

At the conclusion of the activity, participants were given discussion questions for 15 minutes. Topics included how the chatbot impacted their coding process, their comfort with allowing students to use the chatbot, necessary modifications for classroom use, strategies for scaffolding student use, and their general interest in integrating AI tools into their classrooms (Figure \ref{fig:days}). 

 \section{Analysis \& Results}
 We conducted the analysis of the PD sessions data in two phases. The first phase focused on behavior, time stamping and affect and emotion tracking, while the second phase involved hybrid thematic analysis related to discussions and questions.

\subsection {Analyzing Behavior, Interaction, and Emotion}

We conducted a multi-source, time-anchored analysis of each ~50-minute PD session using (a) Zoom video and screen recordings, (b) think-aloud transcripts, and (c) interaction traces (chatbot prompts, revisions, tab switches, and help-seeking). 
Each video was converted into a minute-level timeline, and using full meeting timestamps, we segmented the recordings into task-based intervals such as coding block-based activities, exploring the chatbot, debugging, addressing connectivity issues, and asking the instructor for help. These segments informed a structured codebook that included activity name, slide reference, timestamp, task description, duration. It also included analytic categories such as emotional responses, comprehension of instructions, interaction patterns with the chatbot, struggles with the interface, and question-asking. Two coders, trained by the lead author, collaboratively coded an initial sample and then independently coded the remaining sessions (each coder responsible for half). After initial annotation, the coders reviewed each session to verify alignment between timestamps, activity labels, and interpretations. Finally, the lead author finalized the tagging for consistency.
To ensure reliability, we applied data triangulation by cross-verifying emotional codes across think-aloud transcripts and screen recordings. We first coded emotional expressions in the transcripts based on verbal cues and then compared these with corresponding actions observed in the screen recordings (e.g., hesitation or task changes). Any discrepancies were resolved through discussion among the research team, ensuring consistency and enhancing the reliability of our findings.
Triangulation enhances validity by confirming consistency across multiple evidence sources \cite{thurmond2001point} and aligns with recent CHI practices that integrate multimodal data streams for more rigorous interpretation \cite{luna2025exploring}. During the transcript analysis, the lead author audited all codes against the transcripts and video, resolved discrepancies, and refined the final labels. 

\subsubsection{\textbf{Timestamp Analysis}}
Time-stamped annotations captured the duration of completing each task and activity (getting familiar with the chatbot and different tabs, try prompt entry and run code in different wave activities), each behavior and feelings such as their moments of hesitation (silence/hovering), prompt responses and revisions, and whether participants sought help from the facilitator or referred to the written instructions. 
Timestamps were treated as bounded intervals (e.g., 24:00–27:00 for question-asking), allowing traceability between behavioral observations and original recordings.

\subsubsection{\textbf{Sentiment Categories}}
From the combined evidence, we identified three sentiment classes expressed during tasks: Positive, Neutral, and Negative. These classes were further divided into fine-grained emotion tags, such as "unsatisfied" (negative) for emotions like annoyed, sad, or upset, and "content" (neutral) for emotions like content, calm, or nonchalant.

\begin{itemize}
    \item Positive: Excited, Confident, Curious, Exploring
    \item Neutral: Satisfied, Content, Bored
    \item Negative: Unsatisfied, Confused, Frustrated
\end{itemize}
 We identified key moments that influenced participants’ emotions and categorized the issues into three analytical lenses: (L1) UI/Interaction (system-level interaction issues), (L2) CT/Science (cognitive CT/science challenges), and (L3) (instruction-following challenges). These lenses helped distinguish the different sources of teachers’ emotions: some arose from breakdowns in system-level interaction (L1), whereas others stemmed from teachers’ cognitive engagement with CT and scientific reasoning (L2) or from challenges in following stepwise instructional guidance (L3). We elaborate on these distinctions in the Discussion.
Examples summarized in the Table ~\ref{tab:commands2} for positive emotions, Table ~\ref{tab:commands3} for Neutral emotions and Table ~\ref{tab:commands4} for Negative emotions.

\begin{enumerate}
    \item Positive:
    \begin{itemize}
        \item \textbf{Excited (L1, L2):} Emerged when they expressed enthusiasm for chatbot functionality and responsiveness and participants could quickly try multiple variants and see immediate, meaningful outcomes (comparing “multiple waves’ visual behavior”, “different prompt tabs”).
        \item \textbf{Confident (L3):} Appeared when task framing was clear and participants anticipated the correct output and navigated options without hesitation.
        \item \textbf{Curious / Exploring (L1, L2, L3): }Marked by prompt tinkering, tab switching to inspect features, and trial-and-error debugging without facilitator help.
    \end{itemize}

    \item Neutral:
    \begin{itemize}
        \item \textbf{Satisfied (L3):} Reported when steps were clear and tasks executed smoothly. Participants progressed but did not express strong feelings and affect.
        \item \textbf{Content (L3):} Participants understood the interface and read/followed instructions together, working steadily without notable difficulty.
        \item \textbf{Bored (L1, L2):} Observed during over-scaffolded or too-simple tasks, long pauses/low interaction, or when UX adjustments (e.g., resizing) felt irrelevant to their goals.
    \end{itemize}

    \item Negative:
    \begin{itemize}
        \item \textbf{Unsatisfied (L1):} Unsatisfiability of the interface prompts disappearing on tab switch, the chat box occupying screen real estate, or difficulty resizing the window.
        \item \textbf{Confused (L1, L2):} Getting confused after several attempts, surfaced with ambiguous onboarding and task-framing and unclear system expectations. (L1: input mechanics — e.g., pressing Enter did not send).
        \item \textbf{Frustrated (L1, L2, L3):} Characterized by repeated retries without progress, incomplete activities, reliance on researchers, or the sense that “the chat takes over the screen,” undermining trust in bot-generated code.
    \end{itemize}
\end{enumerate}

\begin{table*}[t]
  \caption{Positive Emotions; Strategies/Behavior and related Quotes. Legend: \ding{169} L1 = User Interface \& system-level interaction; \textbullet ~ L2 = Computational Thinking \& Scientific reasoning; $\rightarrow$ L3 = Instructional guidance.}

  \resizebox{0.95\textwidth}{!}{%
  \begin{tabular}{|p{0.15\textwidth}|p{0.4\textwidth}|p{0.35\textwidth}|}
    \hline
    \makecell[c]{\textbf{Positive} \\ \textbf{Emotion}} &
    \makecell[c]{\textbf{Strategies/Behavior} \\ \textbf{The pair of teachers …}} &
    \textbf{Examples of Quotes} \\
    \hline

    \textbf{Excited} &
      \begin{itemize}[leftmargin=*]
        \item[] \ding{169} Express enthusiasm about the chatbot’s functionality
        \item[] \ding{169} Exhibit excitement by engaging in multiple trials very quickly
      \end{itemize}
      \begin{itemize}[leftmargin=*]
        \item[] \textbullet ~ ~ Discuss how the chatbot would be useful to them or students
        \item[] \textbullet ~ Show marked interest in using the chatbot to promote student learning
        \item[] \textbullet ~ Eagerness in seeing the running code
        \item[] \textbullet ~ Show excitement in testing various features
        \item[] \textbullet ~ Show interest by comparing multiple waves’ visual behavior
      \end{itemize}
    &
      \begin{itemize}[leftmargin=*]
        \item[] \ding{169} ``The chatbot functions look very nice''
      \end{itemize}
      \begin{itemize}[leftmargin=*]
        \item[] \textbullet ~ ``I'm really excited that it worked!''
        \item[] \textbullet ~ ``If we were doing this with kids or teachers, we could use the Coach tab to walk them through the steps''
        \item[] \textbullet ~ ``I like that it helps focus the learning.''
        \item[] \textbullet ~ ``We could add a third sprite so that when you click it, it gets absorbed by the third material.''
      \end{itemize}
    \\
    \hline

    \textbf{Confident} &
      \begin{itemize}[leftmargin=*]
        \item[] $\rightarrow$ Demonstrate a strong understanding of the task
      \end{itemize}
      \begin{itemize}[leftmargin=*]
        \item[] \textbullet ~ See the correct and expected output with confidence
      \end{itemize}
      \begin{itemize}[leftmargin=*]
        \item[] \ding{169} Quickly navigate through options; follow prompts without hesitation
      \end{itemize}
    &
      \begin{itemize}[leftmargin=*]
        \item[] $\rightarrow$ ``What needs to be done here is…''
        \item[] $\rightarrow$ ``I’ve got this..''
      \end{itemize}
      \begin{itemize}[leftmargin=*]
        \item[] \textbullet ~ ``Exactly what I was expecting..''
        \item[] \textbullet ~ ``Clearly light passes through the glass more easily''
      \end{itemize}
      \begin{itemize}[leftmargin=*]
        \item[] \ding{169} ``This should be prompted with the coach tab…''
      \end{itemize}
    \\
    \hline

    \textbf{Curious} &
      \begin{itemize}[leftmargin=*]
        \item[] \ding{169} Interacts and prompts the chatbot
      \end{itemize}
      \begin{itemize}[leftmargin=*]
        \item[] $\rightarrow$ Starts project smoothly; showed curiosity in using instructions and other times trying independently
      \end{itemize}
      \begin{itemize}[leftmargin=*]
        \item[] \textbullet ~ Acts intuitively while exploring the chatbot’s features
        \item[] \textbullet ~ Shows enthusiasm and curiosity in materials
        \item[] \textbullet ~ Shows curiosity by switching between the tabs
        \item[] \textbullet ~ Indicates desire to understand the functionality
      \end{itemize}
    &
      \begin{itemize}[leftmargin=*]
        \item[] \ding{169} ``Loaded and ready, let’s dive in''
        \item[] \ding{169} ``Is it possible to tweak the access?''
      \end{itemize}
      \begin{itemize}[leftmargin=*]
        \item[] $\rightarrow$ ``Like to know how my students use the bot with instructions''
      \end{itemize}
      \begin{itemize}[leftmargin=*]
        \item[] \textbullet ~ ``What’s the difference between Coach and Q\&A here?''
        \item[] \textbullet ~ ``Let's see how this next tab works?''
      \end{itemize}
    \\
    \hline

    \textbf{Exploring} &
      \begin{itemize}[leftmargin=*]
        \item[] \ding{169} Delve into the different functions and buttons
        \item[] \ding{169} Debugs using trial-and-error
      \end{itemize}
      \begin{itemize}[leftmargin=*]
        \item[] \textbullet ~ Engages with the outcome of replacing different materials
        \item[] \textbullet ~ Finds driver-navigator coding interesting
        \item[] \textbullet ~ Tests when conditionals are required
        \item[] \textbullet ~ Struggles with typing out the message in the prompt box and try to solve the challenges
        \item[] \textbullet ~ Compare the effectiveness of different materials
        \item[] \textbullet ~ Trying different examples in prompting and observe the result
        \item[] \textbullet ~ Show more goal oriented behavior
        \item[] \textbullet ~ Debugs by the most experienced person, no help of chatbot (debugging by testing and adjusting the code structure until it worked)
      \end{itemize}
    &
      \begin{itemize}[leftmargin=*]
        \item[] \ding{169} ``I’ll change the prompt and see how response will change?''
        \item[] \ding{169} ``Which tabs have the best response?''
      \end{itemize}
      \begin{itemize}[leftmargin=*]
        \item[] \textbullet ~ ``I swap mirror for glass, does the wave still reflect?''
        \item[] \textbullet ~ ``If I flip the direction and it reverses, we’ve found the issue''
        \item[] \textbullet ~ ``Is 180 degrees the reverse direction in bot?''
      \end{itemize}
    \\
    \hline

  \end{tabular}}
  \Description{This table presents the relationship between positive emotions, the strategies and behaviors of teachers, and related quotes from the study. Emotions include Excited, Confident, Curious, and Exploring.}
  \label{tab:commands2}
\end{table*}

\begin{table*}[t]
 \caption{Neutral Emotions; Strategies/Behavior and related Quotes. Legend: \ding{169} L1 = User Interface \& system-level interaction; \textbullet ~ L2 = Computational Thinking \& Scientific reasoning; $\rightarrow$ L3 = Instructional guidance.}
   
  \resizebox{0.95\textwidth}{!}{
  \begin{tabular}{|p{0.15\textwidth}|p{0.4\textwidth}|p{0.35\textwidth}|}
    \hline
    \makecell[c]{\textbf{Neutral} \\ \textbf{Emotion}} &
    \makecell[c]{\textbf{Strategies/Behavior} \\ \textbf{The pair of teachers …}} &
    \textbf{Examples of Quotes} \\
    \hline
    
    \textbf{Satisfied} &
    \begin{itemize}[leftmargin=*]
          \item[] $\rightarrow$ Find their steps in Coding are true
    \end{itemize}

    \begin{itemize}[leftmargin=*]
      \item[] \ding{169} Run the code smoothly
        \item[] \ding{169} Code reuse via duplication instead of re-prompting chatbot
    \end{itemize}
    
    \begin{itemize}[leftmargin=*]
      \item[] \textbullet ~ Enjoyed watching the outcome of the chatbot 
    \end{itemize}
&
    \begin{itemize}[leftmargin=*]
        \item[] $\rightarrow$ “This is cool. Look at me go!”
        \item[] $\rightarrow$ “We knew what we were doing, and the chatbot made it even more fun.”
        \item[] $\rightarrow$ “I wanted to figure it out on my own—and I did.
    \end{itemize}
   
    \\
    \hline

    \textbf{Content} &
        \begin{itemize}[leftmargin=*]
        \item[] \textbullet ~ Understands the interface
        \end{itemize}
 
    \begin{itemize}[leftmargin=*]
      \item[] $\rightarrow$ Struggles with the codes without any challenge
      \item[] $\rightarrow$ Reads/ follows the instructions together
    \end{itemize}

    &
\begin{itemize}[leftmargin=*]
       \item[] \textbullet ~  “No issues here, it’s working as expected.”
\end{itemize}
 
 \begin{itemize}[leftmargin=*]
 \item[] $\rightarrow$   “Okay, that clicked right away.”
\item[] $\rightarrow$ “You scroll, I’ll call out the next instruction.”
\end{itemize}
 \\
    \hline
    
    \textbf{Bored} &
    \begin{itemize}[leftmargin=*]
      \item[] \ding{169} low engagement with chatbot
      \item[] \ding{169} Not struggling
      \item[] \ding{169} Long pauses / low interaction
      \item[] \ding{169} Low relevance in UX appearance resizing
    \end{itemize}
\begin{itemize}[leftmargin=*]
    \item[] \textbullet ~ Dragging of blocks with no new idea or logic
\item[] \textbullet ~
\end{itemize}

 \begin{itemize}[leftmargin=*]
    \item[] $\rightarrow$  Find following the instructions tedious
 \end{itemize}
 &
 \begin{itemize}[leftmargin=*]
     \item[] \ding{169} “No need to ask the bot; I’ll just copy.”
      \item[] \ding{169} “I can do it myself.”
 \end{itemize}

 \begin{itemize}[leftmargin=*]
     \item[] $\rightarrow$  “Instructions feels like a chore”
 \end{itemize}
    \\
    \hline
  \end{tabular}}
\Description{This table presents the neutral emotions expressed by teachers and their associated strategies, behaviors, and example quotes during interaction with the chatbot. The table is divided into three sections: Neutral Emotion, Strategies/Behavior, and Examples of Quotes.
Satisfied: Teachers felt satisfied when their code ran smoothly and when they successfully reused code without needing to re-prompt the chatbot. They also enjoyed watching the outcome of the chatbot's response. Example quotes include: "This is cool. Look at me go!" and "I wanted to figure it out on my own—and I did."
Content: Teachers in this category understood the interface and followed the instructions without difficulty, though they sometimes struggled with the code. Example quotes include: "No issues here, it's working as expected" and "You scroll, I'll call out the next instruction."
Bored: Teachers displayed low engagement with the chatbot and showed little struggle, but expressed dissatisfaction with the task. They experienced long pauses or low interaction, and the task felt overly simple or tedious. Example quotes include: "I can do it myself" and "Instructions feel like a chore."
The table also includes a legend:
Diamond represents user interface and system-level interaction (L1).
Bullet list represents computational thinking and scientific reasoning (L2).
Arrow list represents instructional guidance (L3).}
  \label{tab:commands3}
\end{table*}


\begin{table*}[t]
  \caption{Negative Emotions; Strategies/Behavior and related Quotes. Legend: \ding{169} L1 = User Interface \& system-level interaction; \textbullet ~ L2 = Computational Thinking \& Scientific reasoning; $\rightarrow$ L3 = Instructional guidance}

  \resizebox{0.95\textwidth}{!}{
  \begin{tabular}{|m{0.13\textwidth}|p{0.4\textwidth}|p{0.35\textwidth}|}
    \hline
    \makecell[c]{\textbf{Negative} \\ \textbf{Emotion}} &
    \makecell[c]{\textbf{Strategies/Behavior} \\ \textbf{The pair of teachers …}} &
    \textbf{Examples of Quotes} \\
    \hline
    
    \textbf{Unsatisfied} &
    \begin{itemize}[leftmargin=*]
          \item[] $\rightarrow$ Trouble finding code blocks
  \item[] $\rightarrow$ Need a minimize button for chatbot
  \item[] $\rightarrow$ Chatbot takes too much space; need to hide it
  \item[] $\rightarrow$ Interface problem: Struggling to resize chat window
  \end{itemize}
&
    \begin{itemize}[leftmargin=*]
    \item[] $\rightarrow$ “Where did the prompt go?, oh it was in other tab”
    \item[] $\rightarrow$ "It was not a good experience I won't use it"
    \item[] $\rightarrow$ “I can’t resize the chatbox height"
    \end{itemize}
   
    \\
    \hline

    \textbf{Confused} &
        \begin{itemize}[leftmargin=*]
        \item[] \textbullet ~ Not understanding unexpected results
        \item[] \textbullet ~ Unclarity in the expected chatbot response
        \end{itemize}
 
 \begin{itemize}[leftmargin=*]
\item[] \ding{169} Prompt disappears when switching tabs
\item[] \ding{169} Pressed Enter; message didn’t send (must click Send)
\item[] \ding{169} Confusion on onboarding and task-framing (Didn’t know the goal (“get to know the 
\item[] \ding{169} chatbot” vs wave activity)
\item[] \ding{169} Unclarity in finding what the issue is after several attempts
Unsure whether costume saved automatically
 \end{itemize}

    &
\begin{itemize}[leftmargin=*]
\item[] \textbullet ~  “I bet the wave is supposed to stop at the mirror… is that what’s happening?”
\item[] \textbullet ~  “What happens if I pull …does it still work?”
\end{itemize}
 
\begin{itemize}[leftmargin=*]
 \item[] \ding{169} “Where did the prompt go?, oh it was in other tab”
\item[] \ding{169} “Is there more text below? How do I scroll the Coach tab?”
\item[] \ding{169} “Can I ask the chatbot where to find the block and still figure it out myself?”
\item[] \ding{169} “Is this pseudocode for non-blocks too, or what am I supposed to follow?”
 \end{itemize}
 
 \\
    \hline
    
    \textbf{Frustrated} &
 \begin{itemize}[leftmargin=*]
 \item[] \textbullet ~ Repeated retries without progress multiple time
\item[] \textbullet ~ Left coding activity incomplete?
 \end{itemize}

\begin{itemize}[leftmargin=*]
\item[] \ding{169}   Enter didn’t send the prompt
\item[] \ding{169} Reliance on researchers, undermining chatbot trust
\item[] \ding{169} Couldn’t find needed blocks
\item[] \ding{169} Teachers disliked chatbot code
\item[] \ding{169} Lowered perceived usefulness of the chatbot/code for classrooms
\item[] \ding{169} Chatbot UI got in the way, or generated code felt unusable
\end{itemize}

 \begin{itemize}[leftmargin=*]
    \item[] $\rightarrow$ Following instructions and getting back was tedious
 \end{itemize}
 &
 
 \begin{itemize}[leftmargin=*]
 
\item[] \ding{169} "keep dragging this chat box out of the way—why won’t it stay put? I can’t even find the new code”
\item[] \ding{169} “This is frustrating: the chat takes over the screen, the scripts are all over the place, and I’m hunting for code that should be right there.
 \end{itemize}

 \begin{itemize}[leftmargin=*]
     \item[] $\rightarrow$  “If a sixth grader saw this, they wouldn’t understand it, and honestly, with the chat in 
     
     .way and no blocks showing, neither do I.”

 \end{itemize}
    \\
    \hline

   \end{tabular}}
\Description{This table presents the negative emotions expressed by teachers, along with their corresponding strategies, behaviors, and example quotes. The table is divided into three sections: Negative Emotion, Strategies/Behavior, and Examples of Quotes.
Unsatisfied: Teachers felt unsatisfied when they had trouble finding code blocks, needed a minimize button for the chatbot, or struggled with the interface (e.g., resizing the chatbot window). Example quotes include: "Where did the prompt go, oh it was in another tab" and "It was not a good experience I won’t use it."
Confused: Teachers exhibited confusion when they didn’t understand unexpected results or when the chatbot didn’t respond as expected. Issues included unclear instructions, disappearing prompts, and confusion during onboarding. Example quotes include: "What happens if I pull... does it still work?" and "Where did the prompt go, oh it was in another tab."
Frustrated: Teachers experienced frustration due to repeated retries without progress, difficulty finding blocks, and dissatisfaction with the chatbot. They also felt undermined by reliance on researchers and had issues with chatbots in classrooms. Example quotes include: "Keep dragging this chat box out of the way—why won’t it stay put? I can’t even find the new code" and "This is frustrating: the chat takes over the screen, the scripts are all over the place, and I’m hunting for code that should be right there."
The table also includes a legend:
Diamond represents user interface and system-level interaction (L1).
Bullet list represents computational thinking and scientific reasoning (L2).
Arrow list represents instructional guidance (L3).}
  \label{tab:commands4}
\end{table*}


\begin{figure*}[t]
  \centering
  \includegraphics[width=\textwidth]{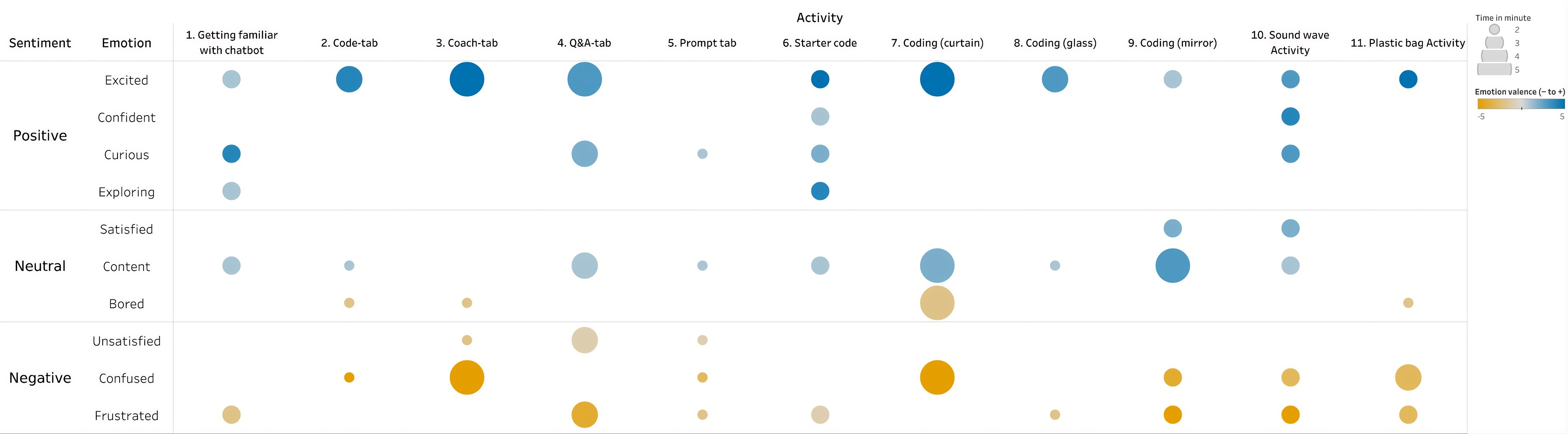}
  \caption{Bubbles encode the number of coded utterances (size); color shows valence (positive/negative/neutral). Large negative bubbles suggest intervention points.}
  \Description{This figure shows a bubble chart where each bubble represents the frequency of coded emotional expressions during various activities. The x-axis lists 11 activities, such as getting familiar with the chatbot, coding tasks, and a plastic bag activity. The y-axis categorizes emotions into three sentiment types: Positive, Neutral, and Negative. Each emotion within these categories (e.g., Excited, Confused, Satisfied) is represented by bubbles that vary in size, indicating the number of coded utterances. The color scale ranges from negative emotions to positive emotions, with neutral emotions in between. Larger negative bubbles suggest potential intervention points. The bubble size also encodes the amount of time spent on each activity.}
  \label{fig:tableau}
\end{figure*}

\begin{figure*}[t]
  \centering
  \includegraphics[width=\textwidth]{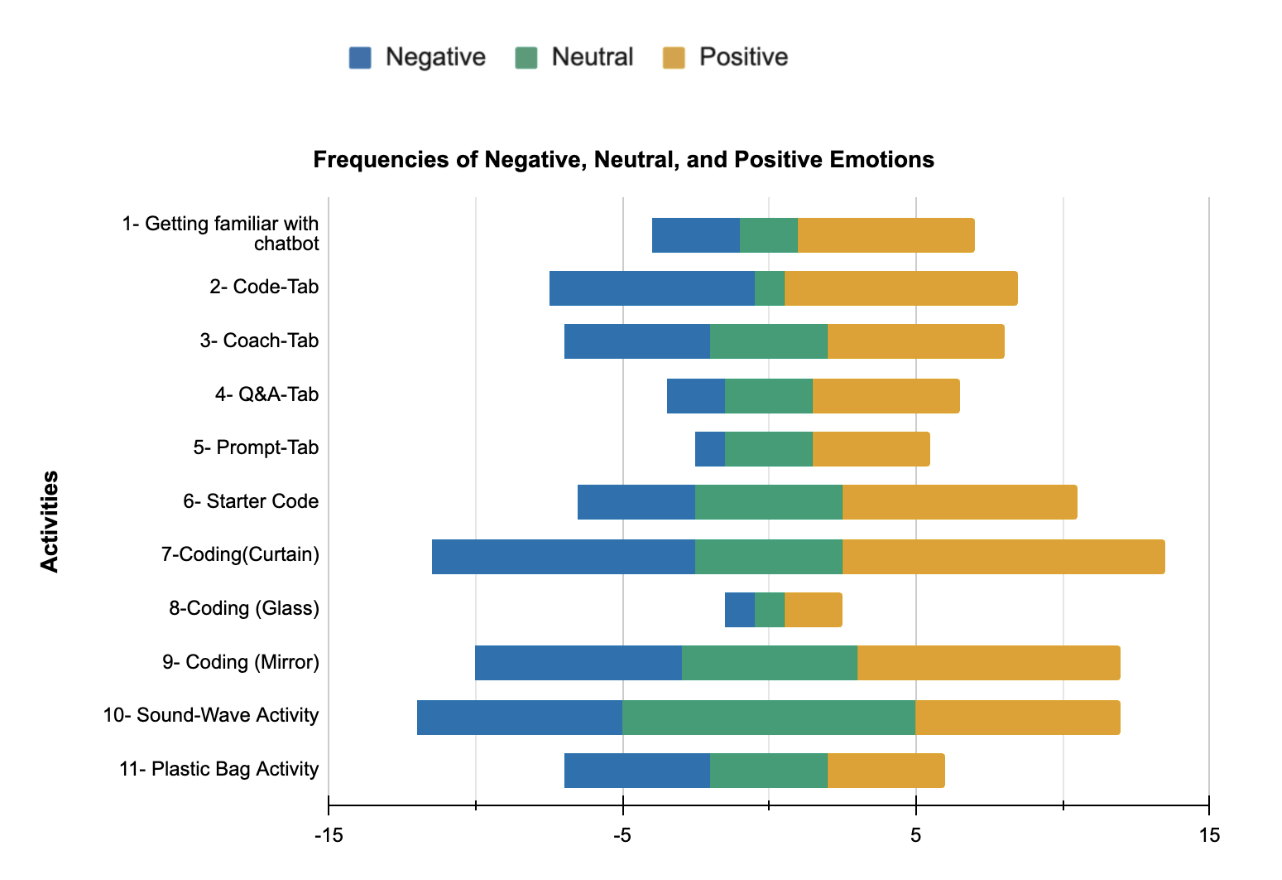}
  \caption{Counts of coded utterances per stage; negative (left), neutral (centered), positive (right), enabling comparison of affect balance across activities.}
  \label{fig:stacked}
  \Description{This bar chart presents the frequencies of negative, neutral, and positive emotions across different activities during the study. The horizontal axis represents frequency, with negative values indicating negative emotions, zero representing neutral emotions, and positive values representing positive emotions. The vertical axis lists the activities. Each activity includes bars for negative, neutral, and positive emotion frequencies.}
\end{figure*}

\subsubsection{\textbf{Affective Responses Across Activities}}
Figure \ref{fig:tableau} visualizes per-activity emotions as bubbles (bubble size : minutes; color : valence from negative to positive).
The chart shows excitement as the most prominent positive emotion, concentrated in Coach-tab, Coding (curtain), with a smaller peak at Plastic-bag activity, while curiosity is most visible in the initial Getting familiar with the chatbot step. Within the neutral emotion, content is most apparent during Coding (mirror). On the negative side, confusion dominates, appearing in Coach-tab, Coding (curtain), Coding (mirror), and the Sound-wave activity. 

Time allocation follows a similar pattern: the largest bubbles occur which indicate the most time spent occurred at Coach-tab (for both positive and negative affect) and Coding (curtain), whereas Starter code and several Sound-wave and Coding (mirror) instances are marked by small bubbles, indicating brief engagement. Moreover, six activities show mostly small bubbles paired with neutral and negative sentiment, suggesting short time involvement of the participants.
Several activities show mixed color within the same column, suggesting within-step fluctuation between positive and negative states. Neutral emotion appears where progress is steady but effect is weak or muted.
The average time spent on each activity was derived by calculating the mean duration across all groups, reflecting the emotional engagement during each task.
Figure \ref{fig:stacked} shows, for each activity, the counts of affect-coded utterances—negative plotted to the left, neutral located at the center, and positive to the right. Segment length indicates frequency (how often the affect occurred), not intensity. Coding (curtain) has the largest negative segment alongside substantial positive, marking it as the most challenging yet engaging stage; this aligns with Exploring groups, who probed features (positive) while encountering friction (negative). Coding (mirror) and Sound Wave skew positive, indicating smoother progress once participants were oriented. Coding (glass) shows low totals across all valences, suggesting limited verbalization/engagement. Early navigation stages (Code-, Coach-, Q\&A-tabs) are net positive with modest negativity, consistent with low-stakes exploration. Overall, the pattern matches our persona analysis: Task-Focused groups advanced efficiently with mostly positive affect, while Exploring groups produced more utterances. 

\subsection{Learner Persona}

Personas are widely used in HCI to identify meaningful user groups by capturing differences in goals, behaviors, and motivations \cite{cooper1999inmates}. Recent AI research emphasizes that personas are especially important in intelligent-system contexts because users’ attitudes toward AI, such as confidence, trust, and willingness to engage directly shape their interaction behavior \cite{holzinger2022personas}. Following this perspective, our personas represent distinct behavioral and emotional orientations toward the AI-enabled chatbot, making them appropriate for interpreting teacher–AI interaction patterns.

\textbf{Persona = (Time-on-task, Emotion trajectory, Strategy pattern, Help Seeking)}

Based on this definition, we identified three personas: 
\begin{enumerate}
    \item Explorer/Task-Focused
    \item Mixed 
    \item Frustrated
\end{enumerate}

Table \ref{tab:personas} summarizes persona-level differences in affect trajectories, strategies, and help seeking.

To construct these personas, we aggregated the coded indicators from our multi-modal dataset help-seeking behavior, task strategies, time on task, and emotional valence into an integrated behavioral–affective profile for each of the 11 groups. This profile summarized both how each group worked and how they felt during the activity. 


\begin{table*}[t]
\caption{Observable Personas, Time Spent, Affect Trajectories, Strategies, and Help Seeking}
\footnotesize              
\setlength{\tabcolsep}{2pt}
\renewcommand{\arraystretch}{1.8} 
\begin{tabular}{|p{0.1\textwidth}|p{0.06\textwidth}|m{0.09\textwidth}|p{0.20\textwidth}|p{0.30\textwidth}|p{0.16\textwidth}|}
\hline
\textbf{Persona} & \textbf{\# of Groups} & \textbf{Time Spent} &
\textbf{Emotion Trajectory} & \textbf{Strategy/Behavior} & \textbf{Help Seeking} \\
\hline

Exploring
& \centering \textbf {6}
& \centering Long
& \multirow{2}{*}{\centering \makecell[l]{\textbf{Mostly Positive/Neutral}\\(Engaged, Curious, Satisfied)}}
& \multirow{2}{0.24\textwidth}{
    \begin{minipage}[t]{0.5\textwidth}
      \begin{itemize}[leftmargin=*,noitemsep,topsep=0pt]
        \item Instruction-led execution
        \item Tab-switch exploration
        \item Prompt refinement \& output comparison
        \item Few stalls
      \end{itemize}
    \end{minipage}
  }

& \makecell[l]{ Low, after several \\ self-retries}
\\
\cline{1-1}\cline{3-3}
Task-Focused  &  & \centering Short &  &  &  \\
\hline

Mixed
&
\centering \textbf{3}

& \makecell[l]{~Average\\Different \\Time \\  Across stages}
& \makecell[l]{\textbf{Variable} (shifts between\\positive and negative \\ across stages)}

&  \begin{itemize}[leftmargin=*,noitemsep,topsep=0pt]
    \item Strategy shifts within session
    \item Following instructions and exploration
  \end{itemize}
& \makecell[l]{Occasional, to recover\\ from stalls}
\\
\hline

Frustrated
& \centering \textbf{2}
& \makecell[l]{Long dwell\\ around \\ failures}
& \makecell[l]{\textbf{Mostly Negative}\\(Confused, Unsatisfied,\\ Frustrated)}
& \begin{itemize}[leftmargin=*,noitemsep,topsep=0pt]
    \item Repeated retries without progress, backtracking
  \end{itemize}
& \makecell[l]{Early/frequent; Signs of\\ Fatigue or \\ Abandonment}
\\
\hline

\end{tabular}
\Description{This table presents the characteristics of four observable personas, including the number of groups representing each persona, the amount of time spent on activities, their emotional trajectory, strategies and behaviors, and help-seeking behaviors. The table is divided into the following columns: Persona, # of Groups, Time Spent, Emotion Trajectory, Strategy/Behavior, and Help Seeking.
Exploring Persona: Comprising 6 groups, these participants spent a long time on activities and showed mostly positive or neutral emotions, including engagement, curiosity, and satisfaction. They demonstrated an instruction-led approach, explored different tabs, refined prompts, and experienced few stalls. Help seeking was low, typically occurring after several self-retries.
Task-Focused Persona: Representing a short time spent on activities, these participants had a clear focus on the task with minimal distractions. Their emotional trajectory was neutral, and their strategy involved following instructions and refining prompts. They did not seek much help during the task.
Mixed Persona: Comprising 3 groups, this persona showed a varied emotional trajectory, shifting between positive and negative emotions across different stages. Their strategies shifted within the session, and they combined instruction-following with exploration. Help seeking was occasional and often used to recover from stalls.
Frustrated Persona: Representing 2 groups, these participants spent a long time dwelling on failures and had mostly negative emotions, such as confusion, dissatisfaction, and frustration. They experienced repeated retries without progress, often backtracking. Help seeking was early and frequent, with signs of fatigue or abandonment.}
\label{tab:personas}
\end{table*}

\begin{figure*}[t]
  \centering
  \includegraphics[width=\textwidth]{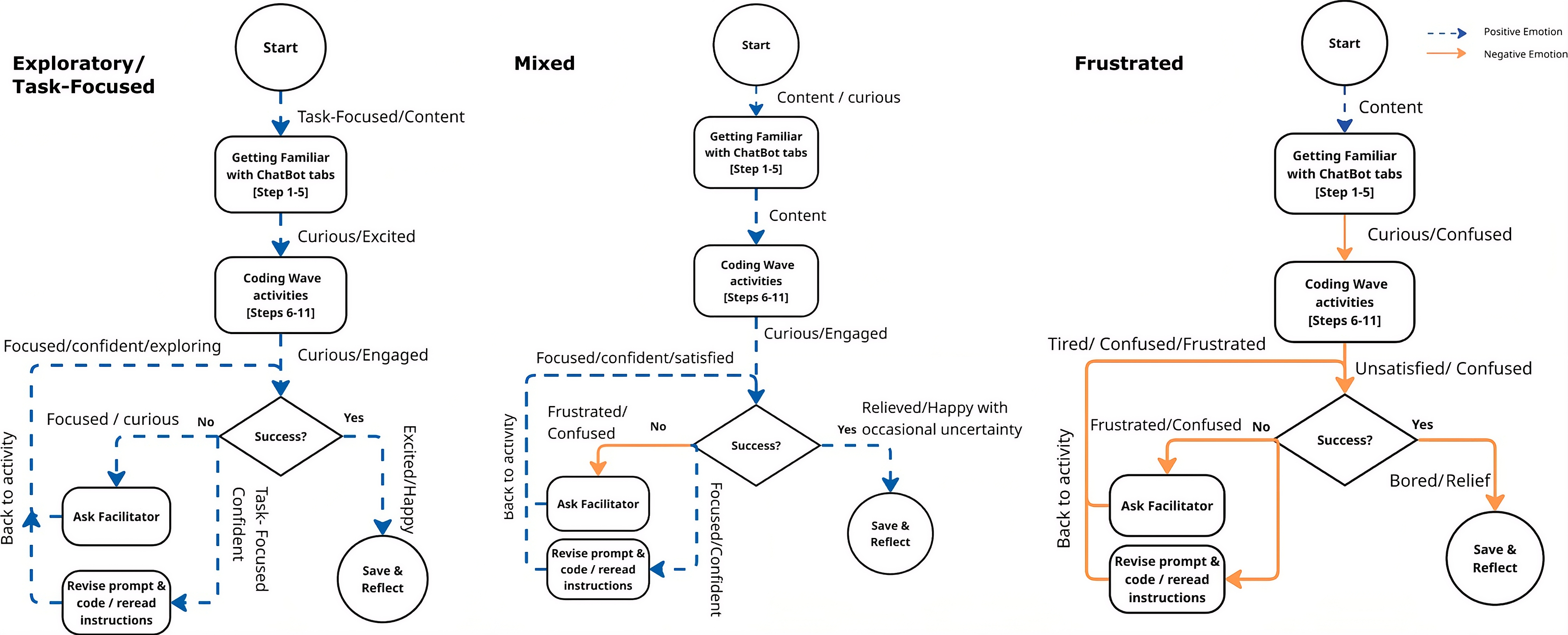} 
  \caption{Different Personas' Reactions and Emotion in Each Stage of Activities}
  \label{fig:flow}
  \Description{This flowchart illustrates the emotional reactions and decision-making pathways of three user personas—Exploratory/Task-Focused, Mixed, and Frustrated—across different stages of the activity. The diagram shows how each persona progresses through the steps of getting familiar with the chatbot and coding wave activities, with emotion and behavior displayed at each stage.
Exploratory/Task-Focused Persona: This persona starts with curiosity and excitement during the initial task (getting familiar with the chatbot), then moves to confidence and engagement during the coding tasks. If they experience difficulty, they may ask the facilitator or revise their prompts. Positive emotions are indicated by dashed blue arrows, and negative emotions by solid orange arrows.
Mixed Persona: This persona shows variable emotions, moving between contentment, curiosity, and frustration. If they encounter issues, they may ask the facilitator or recheck instructions, and if they succeed, they experience relief and happiness. The flow highlights both positive and negative emotional trajectories.
Frustrated Persona: This persona begins with confusion and frustration, particularly during the initial phase of using the chatbot. They experience repeated frustration and struggle, reflected by a strong focus on negative emotions. However, if they succeed, they feel relief and are able to proceed.
Arrows represent the decision-making process and emotional responses at each stage, with blue dashed arrows indicating positive emotions and orange arrows indicating negative emotions.}

\end{figure*}

\begin{figure*}[t]
  \centering
  \includegraphics[width=\textwidth]{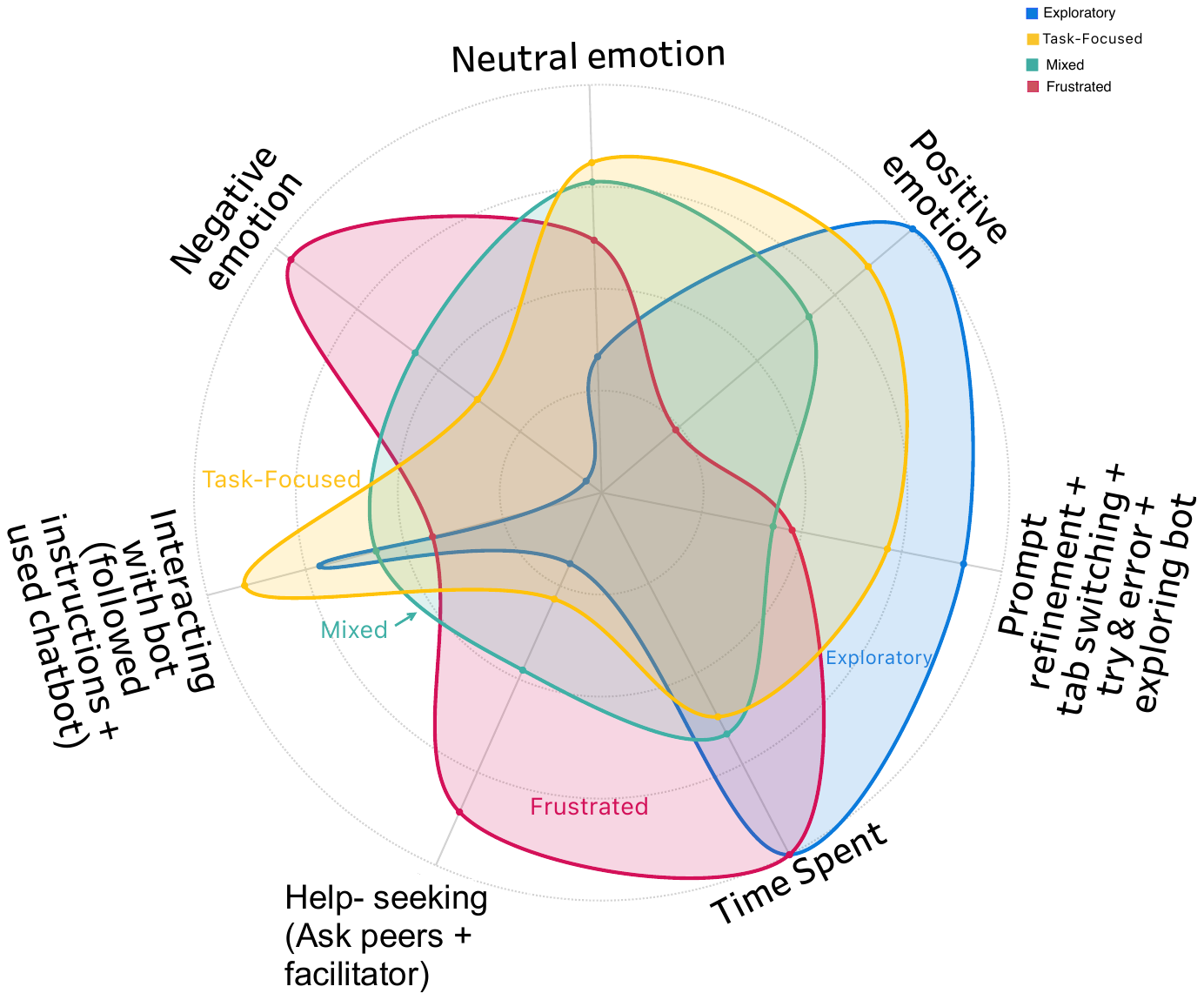}
  \caption{Radar chart comparing four personas—Exploratory, Frustrated, Mixed, and Task-Focused—across seven dimensions: three emotion states (negative, neutral, positive), time spent, help-seeking, instruction-following bot queries, and refinement-oriented interactions (prompt editing, tab switching, and trial-and-error exploration). Higher values indicate more frequent or intense presence of that dimension for a given persona.}
  \label{fig:radar}
  \Description{A radar chart comparing the behavioral and emotional profiles of four user personas: Exploratory, Frustrated, Mixed, and Task-Focused. The personas are mapped across axes representing emotions (Positive, Neutral, Negative) and interaction behaviors (Exploratory, Prompt refinement/Trial \& Error, Time Spent, Help-seeking, and Interacting with bot). Exploratory users peak in Positive Emotion, Exploratory behavior, and Time Spent. Frustrated users spike in Negative Emotion and Help-seeking, with high Time Spent but low Positive Emotion. Task-Focused users peak in Interacting with the bot (following instructions) and Neutral Emotion. Mixed users show a moderate profile across dimensions between the other groups.}
\end{figure*}

Figure \ref{fig:flow}, visualizes time-aligned interaction flows for the three personas (Exploratory/Task-focused, Mixed, Frustrating). Each diagram is built from our time-stamped screen recordings and think-aloud: nodes show blocks of activities, Steps 1–5 (onboarding/ getting familiar with chatbot tabs) and Steps 6–11 (wave activities such as sound and light), plus decision points (e.g., Success?), help-seeking (Ask Facilitator), and outcomes (Save \& Reflect).
Edges encode the temporal path to the activities edge; edge style encodes valence (dashed = positive, solid = negative, neutral segments are explicitly labeled as content). Each edge is annotated with the emotion at that step (e.g., Curious, Confident, Confused, Frustrated). Loops indicate retries/backtracking (e.g., revise prompt \& code / reread instructions), and Start/End are shown explicitly.

We visualized the aggregated coded features using a radar chart (Figure \ref{fig:radar}), plotting the frequency of key behaviors alongside the magnitude of time spent. The chart compares the four personas across seven dimensions: positive, neutral, and negative emotions, time spent, help-seeking, instruction-following queries, and refinement behaviors (e.g., prompt editing, tab switching, trial-and-error). 

This visualization revealed clear clustering across groups and served as the analytic basis for our personas, illustrating how each group’s emotional and behavioral patterns diverge in meaningful ways.

Interpretation of the Radar charts based on each persona:
\paragraph{Exploratory.}
Exploratory groups (1, 3, 7, 11) spent the longest time on task, showed the highest positive emotion, and frequently refined prompts, switched tabs, and experimented with bot features. They needed little help and demonstrated curiosity-driven experimentation.  

\paragraph{Task-Focused.}
Task-Focused groups (2,8) showed high positive and neutral emotion with minimal negative affect. They followed instructions efficiently, interacted with the chatbot when needed, and completed tasks quickly with little help-seeking. Exploratory and Task-focused personas differed primarily in efficiency versus depth of exploration; we therefore report them together.

\paragraph{Mixed.}
Mixed groups (4, 5, 6) displayed moderate levels of positive, neutral, and negative emotion. Their engagement fluctuated across stages: they alternated between following instructions, brief exploration, and occasional frustration. They sought help periodically when stuck but generally continued independently.

\paragraph{Frustrated.}
Frustrated groups (9, 10) showed the highest negative emotion and the lowest positive emotion. Although they spent substantial time on the activity, much of it was directed toward repeated retries and heavy help-seeking. Their interactions reflected confusion, stalled progress, and difficulty navigating both the bot and the task.

\subsection{Thematic analysis of Open-Discussion Data: Evaluating Perceptions and Planned Pedagogical Practices}
In the second phase of the analysis, we conducted a thematic analysis of the post-question and discussion responses using a combined deductive and inductive approach \cite{fereday2006demonstrating}. We created deductive parent themes from the discussion questions and answers (confidence, teacher preparation, student use, required changes, scaffold changes) and inductive sub-themes emerged within those domains. Four researchers co-developed the themes and the codebook: Two trained researchers independently coded the open-discussion questions. Two additional researchers re-reviewed and refined the codes, grouping them into sub-themes and higher-order themes. They resolved the disagreements through discussion to agreement, and finalized the thematic structure and used them in the results.
Our analysis resulted in 144 codes which two researchers sorted into 20 categories which can be found in Table \ref{tab:themes}. 
Those categories were then sorted into 6 themes: 1) Benefits \& risks to students, 2) pedagogical practices around chatbot, 3) assisting all learners, 4) chatbot allowance in classrooms, 5) benefits for teachers, and 6) usability.

\begin{table*}[t]
\caption{Themes and categories produced from thematic analysis of teacher discussion groups.}

\centering
\setlength{\tabcolsep}{5pt}
\renewcommand{\arraystretch}{1.15}

\begin{tabularx}{0.92\textwidth}{|
  >{\RaggedRight\arraybackslash}p{0.36\textwidth}|
  >{\RaggedRight\arraybackslash}X|}
\hline
\textbf{Themes} & \textbf{Categories} \\
\hline

\multirow{4}{=}{Benefits \& Risks to students} &
Chatbot may hurt learning to program \\
& Building student prompting skills \\
& Chatbot can increase student self-efficacy \\
& Chatbot can build student knowledge \\
\hline

\multirow{5}{=}{Pedagogical Practices Around Chatbot} &
Build foundational learning first \\
& Scaffold intro to chatbot \\
& Control of chatbot features \\
& Chatbot and pair programming \\
& Feature requests for engagement \\
\hline

\multirow{5}{=}{Assisting All Learners} &
Chatbot can support novice students \\
& Chatbot can provide challenge to advanced learners \\
& Chatbot not appropriate for advanced learners \\
& Chatbot can help with differentiation \\
& Feature requests for accessibility \\
\hline

\multirow{2}{=}{Allow chatbot in classroom?} &
Would allow students to use the chatbot \\
& Would not allow students to use the chatbot \\
\hline

\multirow{3}{=}{Benefits for teachers} &
Chatbot builds teacher confidence \\
& Chatbot can support teacher instruction \\
& Chatbot can support student independence \\
\hline

Usability & Usability problems \\
\hline
\end{tabularx}
\Description{This table presents the themes and categories derived from the thematic analysis of teacher discussion groups. The table is divided into two columns: Themes and Categories.
Benefits \& Risks to Students: Categories include how the chatbot might hurt students’ learning, such as potentially hindering the development of programming skills, and its potential benefits, such as increasing student self-efficacy and building knowledge.
Pedagogical Practices Around Chatbot: Categories here explore teaching practices related to the chatbot, such as building foundational learning before introducing the chatbot, scaffolding its use, controlling its features, and using pair programming alongside the chatbot. It also includes teacher requests for better engagement features.
Assisting All Learners: Categories focus on how the chatbot can support different learners, including novice students, those needing challenge, and how the chatbot can assist in differentiation. There are also feature requests for improving accessibility.
Allow Chatbot in Classroom?: The table outlines teachers' positions on allowing students to use the chatbot in the classroom, with some wanting to allow its use and others not.
Benefits for Teachers: Categories include the chatbot’s role in building teacher confidence, supporting teacher instruction, and aiding student independence.
Usability: This category identifies usability issues with the chatbot, highlighting problems that need to be addressed for a better user experience.}
\label{tab:themes}
\end{table*}

\subsubsection{\textbf{Benefits to students}}
Teachers recognized that the chatbot could benefit students in a number of ways. First the chatbot could increase student self-efficacy. A teacher in Group 10 mentioned that some students lack self belief around programming and with the chatbot students may realize “hey, I can go to town on stuff like this... They may just really feel much more confident with the chat bot”, meaning that students might be able to work faster and then feel better about their abilities. Additionally, they identified that the chatbot could help students build their computational thinking (CT) skills. For example, the chatbot can build students’ CT vocabulary; Group 2 said that the chatbot could “reinforce that vocabulary, and then… you know, and after they see it and hear it, they start. They start speaking it.” 

Six of the groups mentioned that the chatbot could help grow their students’ prompting skills. Group 1 mentioned that “I feel like I would use it because I would want my students to like, learn what prompts do you need to use and like get familiar with [prompting]”. Groups 1 \& 10 mentioned the balance between teaching programming basics and practical skills. 

\subsubsection{\textbf{Risks to students}}
Group 10 mentioned that the chatbot changed the learning focus from “How do I use block coding” to “learning about, how do I write this prompt to get me to what I want?”. Group 10 said that they do not mind the shift towards prompting, as some skills fade in importance, “coding is like becoming cursive, well now we type”. Some of the groups were concerned about this shift. Group 6 reflected on the loss of productive struggle that students might lose out on, “I didn't feel like I was making headway in the same way that I do when I'm just messing around with blocks on my own, you know. Like, I felt like it was frustration that wasn't productive, because I couldn't figure out what it wanted me to do”. Teachers from Group 4 expressed concern that students might not truly understand the code they generated. They worried that because the chatbot "gives them exactly what to do," students would just copy the code without understanding, questioning, "did they actually understand what was happening? Or did they just copy that?"

\subsubsection{\textbf{Pedagogical Practices Around Chatbot}}
All groups mentioned that they would scaffold student introduction to the chatbot through guided examples, a prompting worksheet, a slidedeck, a video tutorial, and explaining the tabs. A teacher from Group 6 highlighted the importance of establishing the chatbot's fundamental purpose, especially for students who have little experience with such tools. The teacher would frame the chatbot as "basically like a tutorial or a help," because many students "don't know that you use chatbots for all different things."

To ensure students still learned foundational skills and did not become overly reliant on the chatbot, teachers discussed several pedagogical safeguards they would implement. Several teachers expressed a desire to build students' foundational programming understanding before introducing them to the chatbot. A teacher from Group 8 felt that giving students the chatbot immediately would be "handicapping" them, likening it to "giving them a textbook that has all the answers." Instead, this teacher would prefer students to "just play" with programming first and then perhaps "have them use the chat on a bigger project." Similarly, a teacher from Group 2 felt that "trial and error is the best way versus giving [the chatbot] to them first." Teachers were concerned that students might use the chatbot without understanding the code it generated. To mitigate this, a teacher from Group 7 would ask students questions to evaluate their comprehension of the code, such as, "How did you do that?" or "So what did you do when the bot made the code for you? What did you do to make sure you understood what the code is?" To prevent students from becoming distracted or cognitively overloaded, teachers mentioned the need for clear structure in assignments. A teacher from Group 3 noted that "kids like structure" and that allowing them to "tinker" without clear guidance might cause them to "shy away from... the task at hand” and said they would design activities so they focus on only one learning objective at a time.

Teachers expressed a desire for control over various chatbot features to better manage student use and learning. A primary request from seven of the groups was the ability to disable specific AI features for students. This control would allow teachers to tailor the tool to a particular lesson's goals. For example, a teacher from Group 5 stated it would be helpful if "the code button is off limits" while still allowing students to use the "QA" (questions and answers) function. Teachers also wanted control over the level of detail the chatbot provided in its responses. 
A teacher from Group 9 envisioned a "tiered approach" where the teacher could set the level of response. A Level 1 response, for instance, might "give me the code," whereas a Level 2 response would "prompt me a little bit, so that I understand what it is that I'm trying to do." Some teachers wanted the ability to limit the number of prompts a student could submit. A teacher from Group 1 suggested this as a way to create a challenge, stating that for students who are already comfortable, the teacher might "limit you to… two questions, you get to ask the chat."

Groups 10 \& 3 mentioned that they could use the chatbot to support students in projects in which there is a high amount of student choice. Groups 10 \& 11 also wanted the chatbot to have gamified features and awards to keep students motivated and engaged.

\subsubsection{\textbf{Assisting all learners}}
Teachers recognized the chatbot's potential to help a diverse range of students, from novices to advanced learners, and those with specific accessibility needs. Eight of the groups mentioned that the chatbot could be a powerful tool for differentiation. For example, a teacher from Group 8 felt that the chatbot would allow accelerated students to "move forward and then give them like more challenges" and "letting them explore a little further." However, a teacher from Group 5 also noted that some "of the higher students wouldn't want [the chatbot]," as they "like to figure it out on their own."

Teachers also saw the potential for the chatbot to assist students with specific accessibility needs. Teachers requested text-to-speech options to support different learning styles. A teacher from Group 10 asked if the chatbot could have "read aloud options... Is there a way that you can hear it?" to accommodate students who struggle with reading. Similarly, a teacher from Group 6 highlighted the need for language and reading support, explaining that some students are "Spanish speaking" and need to "hear it in Spanish, or read it in Spanish." This teacher also noted that some students "read at a second grade level" and would find the large amount of text "really overwhelming."

\subsubsection{\textbf{Benefits to teachers}}
Teachers in Groups 1, 2, and 9 felt that the chatbot could help them with leading activities in their classrooms. A teacher in Group 1 said that it would make them feel “more comfortable” working with students because every student’s code is different, and having a chatbot that can interpret student code is like “having an answer key going into an activity which is nice.” Another teacher from Group 9 commented that the chatbot “would make [them] feel really confident about going into a teaching situation, because some kids [ask] like, ‘How do I do this?’ I'm like ‘I have no idea’”. Another teacher in Group 9 felt like the chatbot had limited ability to support them in leading activities because, “It showed what kind of code I needed to use… But, It didn't explain why”.

Five of the groups mentioned that the chatbot could support teachers with student interventions. The chatbot could take "pressure off of the teacher" (Group 7), as each student could get support without the teacher having to help everyone individually. Similarly, a teacher from Group 9 explained that “you have 28 kids, but you're only one person” and that helping each individually isn’t always possible. 

Teachers also saw the chatbot as a way to build student independence, which would free up their own time and energy. By providing students with a tool for debugging and troubleshooting, the chatbot could empower them to solve problems on their own. One teacher from Group 2 noted that the chatbot’s questioning approach was similar to what they would use to guide students, stating, "I appreciated the questions that it asked, because those are questions that I would ask." Another teacher from Group 4 noted that the chatbot would be "handy for troubleshooting," especially when students need help with "little minor details" while a teacher is busy with other students.

\subsubsection{\textbf{Chatbot allowance in classrooms}}
All the groups reported that they would allow their students to use the chatbot, though many had specific ideas about which features to use and when. The teachers' decisions often depended on their primary pedagogical goal: whether they were focused on teaching core computational thinking (CT) skills or using the chatbot as a tool to teach other domain knowledge. Some teachers were less concerned about students learning CT concepts and more focused on using the chatbot as a medium to teach other subjects. For these teachers, using all of the chatbot's features, including the code generation function, was acceptable. A teacher from Group 1 explained that in their classroom, they wouldn't use the chatbot to necessarily teach coding, but rather to use it as a "medium to teach something about like insects." For example, they would have allowed students to use full chatbot functionality on the activity given during the co-design sessions, noting that even with the chatbot, students would still have to "think about that, because the prompt didn't work when you just said to make it."

In contrast, four of the groups (3, 5, 6, and 8) stated they would not allow students to use the code generation tab, at least in the beginning. They felt that this feature would hinder students' ability to learn foundational programming concepts. A teacher from Group 8 felt that the code generation feature "giv[es] you the answer too quickly." This teacher compared the process to "copying and pasting" and felt that it prevented students from engaging in the "cognitive ability of transferring information from one place to another."

 \section{Discussions}
Without prompting, teachers have highlighted a suite of needs for CT-integrated instruction involving block-based programming that have been identified in prior research about novice programming and new educational tools \cite{bosch2017affective}. The theme of pedagogical practices shows that teachers know both that students need to learn foundational skills before being offered shortcuts (like a code generator), but also that students must be taught how to use new tools like the chatbots, and this need is separate from learning the content being taught within the tool. 

The theme of chatbot benefits and risks highlights the possibility of adaptive individual tutoring - since they learn from the chatbot embedded within the environment, students can build knowledge and self-efficacy. Furthermore, teachers identified that similar chatbots will likely be used in students’ future careers, so practicing with prompting chatbots is important learning. The theme of assisting all learners addresses an important need teachers have expressed about learning to integrate new CT curricula, that part of the teacher’s  job is to differentiate the educational materials for students with differing prior preparation and needs. Teachers see the potential for the chatbot to reduce the need for teachers to prepare additional materials or adjustments ahead of time since the chatbot can meet some of the learner's needs for extra explanations, language translations, speech to text, or further challenges. 

The theme of benefits for teachers highlights that teachers could immediately see impacts for themselves – building their own self confidence and knowledge about programming, and reducing the need for them to provide help to every student during programming, while also increasing student independence or ability to find help for themselves.

\subsection{RQ1: Affect \& Attitude}
In this paper, we investigated how interacting with a chatbot for learning programming impacts teachers' affect and attitudes. Our analysis revealed that teachers' emotional responses and attitudes varied depending on their engagement with the chatbot \cite{hlee2023understanding}. 

Teachers showed excitement and curiosity when exploring the chatbot's features, while some felt neutral, such as contentment, when tasks were smooth to follow. However, there were also instances of boredom during repetitive or overly scaffolded tasks, which resulted in reduced engagement. Additionally, they experienced frustration and confusion when facing with interface issues, unexpected or mismatched responses. These emotional fluctuations highlight how teachers' experiences with the chatbot affect their attitudes.

Learners' emotions depend on interactions, and resolving barriers can shift feelings from negative to positive. This highlights the difference between LLM and human support, emphasizing when both are needed. Understanding this helps teachers anticipate and address students' needs in moments of confusion or frustration, whether due to lack of guidance, usability issues, or difficulty understanding. This insight empowers teachers to create a more supportive, adaptive learning environment more effectively.
Our lens coding, (L1) UI/Interaction (system-level interaction issues), (L2) CT/Science (cognitive CT/science challenges), and (L3) Instruction-Following (instruction-following challenges)—clarifies which emotions were primarily driven by interface limitations versus teachers’ own cognitive or instructional work. L1 emotions were tightly coupled to system-level interaction issues, such as disappearing prompts, the chat pane obscuring code, or confusion around input mechanics, all of which reliably produced confusion, unsatisfaction, or frustration even when teachers understood the underlying CT/science concepts. In contrast, L2 and L3 emotions were more often associated with cognitive and pedagogical demands: teachers’ curiosity, confidence, or confusion when reasoning about wave interactions, debugging logic, or interpreting stepwise written instructions, even when the interface behaved as intended. This distinction helps separate emotions that signal breakdowns in the chatbot interface or coding environment (L1) from emotions that indicate productive or unproductive struggle with content or instructional framing (L2/L3).
\subsection{RQ2: Perceptions of LLM use for classroom}
Our findings highlight teachers' perspectives on both the benefits and risks of using large language models (LLMs) in block-based programming education. Teachers noted that LLMs can boost students' confidence and self-efficacy \cite{kumar2024guiding}, helping them make faster progress, developing computational thinking skills, allowing students to build foundational programming knowledge step by step \cite{hobert2023fostering}. Teachers also noted that interacting with LLMs can enhance students' prompting skills, which is an essential skill for engaging with AI-powered tools \cite{sawalha2024analyzing}.
Teachers raised concerns about students over-reliance on chatbot for programming, which could prevent students from understanding programming concepts to simply writing prompts, undermining the value of "productive struggle" in learning. Secondly, they cautioned that chatbots could diminish critical thinking, problem-solving, debugging, and troubleshooting skills. Chatbots also can help teachers by reducing workload, saving time and free teachers to help students with more complex concepts. However, they worried about the lack of transparency behind chatbot-generated code making it difficult to assess students' understanding \cite{yin2025responsible}. Many teachers desired more control over chatbot features, such as limiting automatic code generation to guide, not replace student work. Teachers emphasized careful, scaffolded integration with clear control features to ensure chatbots enhance learning without undermining essential skills development.

One interesting finding in our results is that teachers differed on whether or not the chatbot would be useful for advanced learners, with some teachers feeling that advanced students would not need or like chatbot support, while others felt that advanced learners could use the chatbot to explore beyond a given assignment. These differing perspectives may be rooted in teacher experiences with students. For example, some experienced CT learners reject systems and supports built for novices. Other teachers may have struggled to come up with ideas for experienced CT learners to extend activities designed for novices, and perceive that the chatbot would enable independent exploration without needing teacher help on the programming language or how to code a new idea their students may have. Research has demonstrated that the high variance in prior experience and motivation for students with regard to CT and programming can be a challenge in the classroom, precisely for these reasons \cite{liao2022exploring}.
\subsection{Mapping into Psychological-Educational Frameworks}

To situate an understanding of teachers' behavioral and emotional dynamics within a broader decision-making framework, we turn to psychological frameworks such as Technology Acceptance Models (TAM) and  Theory of Planned Behavior (TPB) which offer complementary lenses for understanding teachers’ attitudes, perceived control, and intentions to use AI systems.

Technology Acceptance Models (TAMs), were first introduced by Davis \cite{davis1985technology, davis1989perceived}, to explain users’ attitudes toward technology through four core constructs: perceived usefulness, perceived ease of use, attitude, and behavioral intention. In educational technology research, TAM has been extended to incorporate pedagogical and learning-related factors \cite{kemp2024testing}. This framework has frequently been used to assess students’ acceptance of AI and to interpret their experiences; for example, studies have shown that students report significantly positive perceptions of ease of use and perceived usefulness when receiving assistance from AI \cite{algerafi2023understanding}.

The Theory of Planned Behavior (TPB) \cite{ajzen1985intentions, ajzen2020theory} provides an additional perspective for examining the factors that shape human actions within teaching and learning environments. TPB explains behavioral intention through attitudes, subjective norms, and perceived behavioral control, which reflects teachers’ confidence in using AI tools in the classroom based on their perceived control over both technical and instructional challenges \cite{ajzen1991theory}. This makes TPB particularly useful for understanding how educators and students decide to engage with technology in educational and research contexts \cite{ivanov2024drivers}.
TPB has been widely used in studies of decision-making processes and has been applied extensively to predict K–12 teachers’ intentions to adopt educational technologies \cite{czerniak1999teachers, salleh2004using}.

Scholars have also employed TPB investigate GenAI adoption in GenAI in education; for example, \citet{ivanov2024drivers} found that perceived benefits of GenAI enhance attitudes and perceived behavioral control, which subsequently increase intention to use these tools.

Researchers such as \citet{abdullah2016developing} have applied TAM-based models to e-learning and found that self-efficacy, social influence, enjoyment, anxiety, and experience significantly shape perceived usefulness and ease of use, while \citet{kemp2024testing} used a similar framework with university students in virtual classrooms and showed that comfort and well-being, cognitive engagement, and access and convenience strongly predict perceived usefulness and behavioural intention to use the technology.
\citet{smolinski2024scaling} demonstrates that LLMs can reliably approximate TAM measures, producing acceptance ratings that align closely with human expert coders. Similarly, TAM has been applied to LLM-driven educational chatbots, showing that students perceive these systems as useful and easy to use and report generally positive attitudes toward adopting them as learning support tools \cite{neumann2024llm}.

Since TPB captures behavioral and control-related factors while TAM focuses on technology-specific perceptions, TPM and TAM can not be used alone for explaining users’ intention to use novel technologies such as AI systems and chatbots. Researchers must integrate the two for a more comprehensive foundation for explaining users’ intentions to adopt AI-driven tools \cite{jiao2024research,mohr2021acceptance}.

\begin{figure*}[t]
  \centering
  \includegraphics[width=\textwidth]{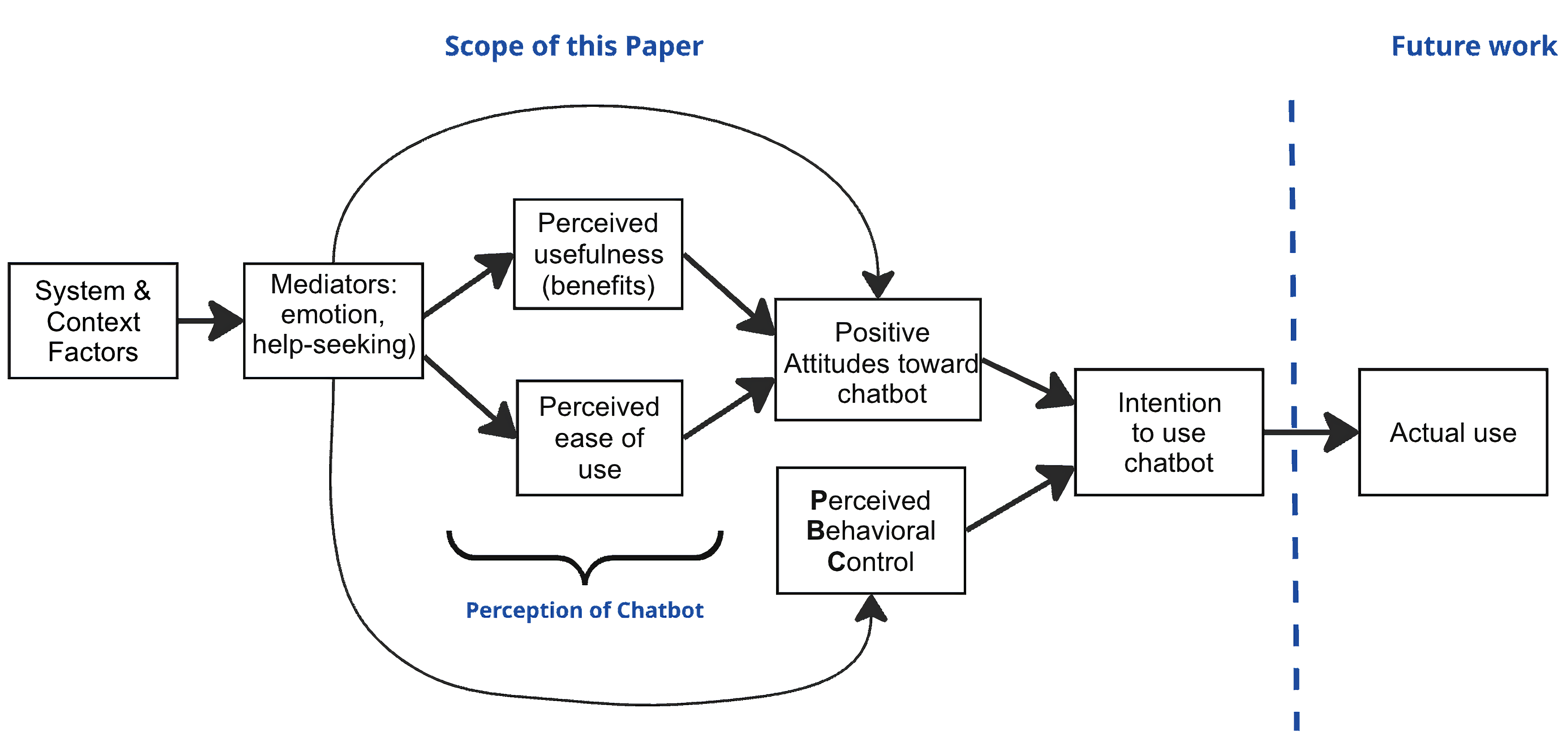}
  \caption{Integrated TAM (Davis, 1989) and TPB (Ajzen, 1991) model used to interpret teachers’ intention to adopt the AI chatbot.}
  \label{fig:framework}
   \Description{A conceptual path model diagram illustrating the scope of the study. The model flows left-to-right, starting with 'System \& Context Factors,' which point to 'Mediators: emotion, help-seeking.' From Mediators, arrows branch out to three destinations: 'Perceived Behavioral Control,' 'Positive Attitudes toward chatbot,' and A bracketed group labeled 'Perception of Chatbot' containing 'Perceived usefulness' and 'Perceived ease of use.' 'Perceived usefulness' and 'Perceived ease of use' also direct arrows to 'Positive Attitudes toward chatbot.' Finally, both 'Positive Attitudes' and 'Perceived Behavioral Control' lead to 'Intention to use chatbot.' A vertical dashed line separates 'Intention to use' from the final box, 'Actual use,' indicating that 'Actual use' is designated as 'Future work,' while the rest of the model represents the 'Scope of this Paper.}
\end{figure*}

In our study, we propose to integrate TAM and TPB to develop a preliminary conceptual foundation for understanding teachers’ interaction, evaluation, and intention to use the chatbot (Figure~\ref{fig:framework}). We conceptualized system and context factors—including the chatbot’s interaction design, the structure of the coding tasks, teachers’ prior experience with block-based programming and AI tools, and the broader pedagogical context, as external variables that may have shaped teachers’ moment-to-moment experiences with the system. Emotions are inseparable from cognition and represent a necessary component of user-centered design \cite{norman2007emotional}; emotional responses can influence engagement and behavior in interactive systems \cite{triberti2017developing}. These encounters elicited affective responses such as confidence, curiosity, frustration, and confusion, alongside teachers’ help-seeking, which may function as mediating processes linking system/context factors to teachers’ evolving perceptions of the chatbot. These mediators may influence teachers’ sense of control, how they navigate difficulties, and the degree to which they feel supported while working with the tool. Based on these mediated experiences, teachers appeared to form preliminary perception about the chatbot’s perceived usefulness—particularly its benefits to students and teachers (Sections 4.3.1, 4.3.5)—and its perceived ease of use.
Prior work suggests that perceived usefulness and enjoyment can contribute to more positive attitudes, which in turn may increase intention to use and continuance intention \cite{jo2023affordance}.

In turn, we hypothesize that these perceptions informed teachers’ attitudes toward the chatbot as well as their perceived behavioral control, reflected in their confidence and self-efficacy when using and managing the tool. 
Together, this model explains our observed patterns in how teacher attitudes and perceived behavioral control (PBC) may have shaped teachers’ intentions to adopt the AI-enabled chatbot in their instructional practice. These relationships collectively motivated our integration of TAM and TPB into a single explanatory framework for modeling teachers’ behavioral intentions (Figure\ref{fig:framework}).

We found that all groups expressed interest in using the chatbot, and we believe that their emotional experiences shaped how and why they intended to use it. For instance, Group 1 viewed the chatbot not for teaching coding, but rather to use it as a medium to teach science and computational thinking, whereas Groups 3, 5, 6, and 8 focused on student learning and preferred limited code-generation features.

These differences align with variation in perceived usefulness, perceived ease of use, and perceived behavioral control, which TAM and TPB identify as key factors shaping emerging intentions to use a technology, and they help explain why each group tends to cluster into distinct personas, as visualized in the radar chart. To make this relationship more concrete, we examine how the radar-chart patterns that define and articulate the personas map onto these theoretical constructs. For example, the Exploratory and Task-Focused profiles—marked by higher positive emotion, active engagement, and lower help-seeking—align with higher perceived usefulness, perceived ease of use, and perceived behavioral control. Their enthusiasm and willingness to try multiple prompts suggest that they see the chatbot as beneficial for achieving their instructional goals (high perceived usefulness); their smooth navigation of features and limited reliance on facilitator support indicate that they experience the system as straightforward to operate (high perceived ease of use); and their confidence in deciding when and how to use the chatbot, as well as their persistence in troubleshooting, reflects a strong sense that they can successfully control and manage the tool in their practice (high perceived behavioral control).

By contrast, the Frustrated profiles—characterized by high negative emotion, frequent requests for help, and repeated breakdowns in task progress—correspond to lower perceived ease of use and weaker behavioral control. Their confusion and dependence on external support signal that interacting with the chatbot feels effortful and unpredictable (low perceived ease of use), and their tendency to abandon or narrow their use of the tool suggests a diminished sense that they can effectively operate it in their own classrooms (low perceived behavioral control), which in turn undermines perceived usefulness. Mixed profiles fall between these extremes: their oscillation between curiosity and frustration, intermittent engagement, and selective help-seeking translate into ambivalent perceptions of usefulness, ease of use, and control. Collectively, the radar chart visualizes how these emotional trajectories and interaction styles may mediate the relationship between system/context factors and teachers’ evolving attitudes and behavioral intentions toward adopting the chatbot.

Our study did not assess actual use of chatbot in classrooms, the findings are consistent with belated frameworks, indicating that emotional states may be playing a central role in technology acceptance, linking engagement with behavioral intention. We posit that these emotions dynamically shaped teachers’ agency, confidence, and decision-making pathways regarding AI integration in educational settings.


In addition to acceptance-based models like TAM, teacher- knowledge frameworks such as Technological Pedagogical Content Knowledge (TPACK) provide a complementary lens for understanding how educators integrate technology into instruction \cite{mishra2006technological}. Research shows that TPACK-informed professional development (PD) helps teachers develop effective pedagogical strategies for integrating GenAI alongside technological skills \cite{shrestha2025tpack}. Their TPACK-aligned PD led to substantial gains in teachers’ perceived ease of use, perceived usefulness, attitudes toward AI tools, intention to use them, actual usage, and self-efficacy—demonstrating the effectiveness of structured, TPACK-guided PD in supporting educators’ adoption of emerging technologies.

 Our findings can also be interpreted through the lens of the TPACK framework. Teachers’ interactions with the chatbot and block-based activities unfolded on top of their existing technological, pedagogical, and content knowledge, meaning that our system and context factors (e.g., chatbot design, task structure, PD scaffolds) were experienced through their evolving TPACK. The affective reactions we observed—such as confidence, curiosity, frustration, confusion, and reliance on facilitator support—highlight moments where teachers were able to mobilize their technological–pedagogical knowledge or, conversely, where gaps in that knowledge became visible. These experiences shaped teachers’ perceived usefulness and perceived ease of use of the chatbot, as well as their attitudes and perceived behavioral control, aligning with prior TPACK-informed PD work that reports gains in ease of use, usefulness, self-efficacy, and intention to use AI tools.

\subsection{\textbf{Design Implications}}
Many negative affective states emerged not because the chatbot produced incorrect content, but because teachers lacked Situation Awareness (SA) of how the system operated. While basic perception (SA Level 1) was sometimes hindered by interface issues like disappearing prompts or hidden code, a deeper breakdown occurred at 'Level 2 SA' (Comprehension). Even when teachers could see the output, they often lacked the transparency to understand its rationale or constraints \cite{jiang2023situation}.

To reduce this confusion, systems should surface clearer explanations of generated outputs. For example, automatically commenting generated code, highlighting specific changes (diffs), and directing users to where code appears can help teachers maintain situational awareness by bridging the gap between raw automation and human understanding. These transparency-oriented design choices address a major gap in teachers’ mental models: knowing what the system did and why it did it (SA Level 2: Comprehension). Establishing this understanding allows teachers to accurately predict how the system will behave under different constraints (SA Level 3: Projection), a necessary step before they can confidently integrate the tool into their teaching practice \cite{sanneman2022situation}.

 When teachers experienced frustration, it often stemmed from stagnation—moments where they tried several prompts but made no observable progress.
These episodes were not limited to interaction and system breakdowns (Analytical Lense L1) ; they also occurred during (Analytical Lense L2) CT/Science (cognitive CT/science challenges) and (Analytical Lense L3) (instruction-following challenges),
when teachers struggled with debugging logic, interpreting scientific explanations, or adapting stepwise guidance. In such situations, simply echoing empathic phrases is not enough to support teachers. Cuadra et al. (2024) showed that conversational agents can express emotional reactions, but struggle with deeper interpretation and follow-up, suggesting that their empathy can feel hollow or performative \cite{cuadra2024illusion}. For teachers, therefore, effective support must go beyond emotional acknowledgment to provide instrumental scaffolding: the system should clarify goals, ask what the user is trying to achieve, or offer to restate the prompt.  Moving beyond performative empathy requires equipping the system with actionable conversational repair strategies—such as offering explicit options or clarifying intent—rather than simply asking users to try again \cite{braggaar2023conversational}. By coupling empathy with these goal-directed repairs, the system shifts from a passive observer to an active partner. This approach not only recognizes the user's emotional state but helps them resolve the underlying breakdown, thereby restoring their agency and competence in the face of automation \cite{jiang2023situation}.

\subsection{Design Recommendations}
\balance
Recommendation 1: Chatbot supporting differentiation in classrooms. Past research has shown that while teachers want to help every student in their classroom one-on-one, they do not have the time to do so, which results in them building pathways for students to independently problem solve \cite{limke2025CHI}. The teachers in our study understood the chatbot to be something that could support differentiation in the classroom while also increasing student independence. To balance student agency and support based on prior skills, we recommend a feature that adapts to students' skill levels, adjustable by teachers or students \cite{huang2024facilitating}. The chatbot could offer more guided prompts initially, decreasing as students improve to encourage autonomy. Additionally, integrating an embedded pre-assessment or sample activity to calibrate prior knowledge and adjust support levels based on performance would be beneficial \cite{chen2023artificial, catete2018infusing}.

Recommendation 2: pedagogical integration of chatbots, allowing teachers to enable or disable the code generation function \cite{almuhanna2024teachers}. This feature would enable teachers to toggle chatbot functionality based on classroom or individualized student needs, such as limiting it to answering conceptual questions or providing code suggestions instead of full code, these controlling features could ensure that students develop problem-solving skills rather than overreliance on chatbots. The teachers in our study feared that code generation would inhibit student learning - especially for students who have not yet mastered basic coding structures.

Recommendation 3: Adapt the chatbot to fit the needs of diverse learners. Our teachers expressed concern over the text-heavy output - especially for students who have lower reading levels or speak English as a second language. Modifying the chatbot to produce visual or auditory output may be another way to support these populations. 

Additional recommendations are to allow users to modify the layout of their environment similar to professional IDEs, as our teachers had strong opinions about the layout which impacted their general attitudes about using the chatbot.

\section{Conclusion}
In this study, we explored how teachers perceive and react to using a Large Language Model (LLM) chatbot in the classroom, with a focus on their emotions and behaviors during interactions. Our key findings highlight the diverse teacher profiles and varying scaffolding needs, illustrating how teachers’ attitudes towards the chatbot can shift based on their experiences. We identified both the benefits and risks of using the chatbot for teaching and learning.
Our study demonstrates that chatbots can provide significant benefits to both teachers and learners. They can accelerate learning and troubleshooting, saving valuable time, and can boost self-confidence, particularly for novice learners. However, different users with varying levels of programming skills benefit from chatbots in different ways. 
Our findings contribute to the ongoing discussion about the use of LLMs in education \cite{kim2025augmented, ravi2025co}, highlighting the importance of fostering critical thinking, ensuring effective interactions, and providing valuable learning experiences \cite{prabhudesai2025here}. We also identified key opportunities for future research, particularly in exploring how AI-driven tools can be designed to prevent over-reliance, avoid replacing teachers’ roles, and mitigate the risk of diminishing critical thinking skills. Additionally, we emphasize the need for design recommendations that maximize the benefits of AI tools while ensuring they meet the diverse needs of users.

\begin{acks}
This material is based upon work supported by the National Science Foundation under Award No. 2405854 and 2405855. Any opinions, findings and conclusions, or recommendations expressed in this material are those of the authors and do not necessarily reflect the views of the sponsor.
\end{acks}

\bibliographystyle{ACM-Reference-Format}
\bibliography{references}

\end{document}